# Origin and stability of the dipolar response in a family of tetragonal tungsten bronze relaxors

Andrei Rotaru[1], Donna C. Arnold[1], Aziz Daoud-Aladine[2], Finlay D. Morrison[1*]

*1 EaStCHEM Research School of Chemistry, University of St Andrews, North Haugh, St Andrews, Fife, KY16 9SH, UK*

*2 ISIS Facility, Rutherford Appleton Laboratory, Chilton, Didcot, OX11 0QX, UK*

**Corresponding Author: e-mail finlay.morrison@st-andrews.ac.uk*

*tel: 01334 463855, fax: 01334 463808*

## Abstract

A new family of relaxor dielectrics with the tetragonal tungsten bronze structure (nominal composition $Ba_6M^{3+}Nb_9O_{30}$, $M^{3+}$ = Ga, Sc or In) were studied using dielectric spectroscopy to probe the dynamic dipole response and correlate this with the crystal structure as determined from powder neutron diffraction. Independent analyses of real and imaginary parts of the complex dielectric function were used to determine characteristic temperature parameters, $T_{VF}$, and $T_{UDR}$, respectively. In each composition both these temperatures correlated with the temperature of maximum crystallographic strain, $T_{c/a}$ determined from diffraction data. The overall behaviour is consistent with dipole freezing and the data indicate that the dipole stability increases with increasing $M^{3+}$ cation size as a result of increased tetragonality of the unit cell. Crystallographic data suggests that these materials are uniaxial relaxors with the dipole moment predominantly restricted to the B1 cation site in the structure. Possible origins of the relaxor behaviour are discussed.

## I Introduction

In recent years there has been a resurgence in research dedicated to discovery of novel ferroelectric materials primarily due to the combination of the requirement of Pb-free alternatives to $PbZr_xTi_{1-x}O_3$ (PZT)[1] and the recent renaissance in multiferroics.[2] To date much of the search for new materials has largely centred on perovskite ($ABO_3$) based materials such as $K_{1-x}Na_xNbO_3$ (as a Pb-free piezoelectric)[3] and $BiFeO_3$ (the most well-known room temperature multiferroic)[4] due to both their compositional flexibility and also our level of understanding of mechanisms to tune properties in this structure type; simple arguments based on steric considerations (tolerance factor), cation and charge ordering and octahedral tilting are all tools to be exploited in the quest for new perovskite ferroelectric and multiferroic materials. More recently an emerging class of materials, tetragonal tungsten bronzes (TTB), have begun to garner renewed interest in the research community. These materials are known to exhibit diverse properties as a result of compositional flexibility, however, whilst ferroelectric TTBs (including $Ba_2NaNb_5O_{15}$[5-7] and $(Ba,Sr)Nb_2O_6$[8-10]) were widely studied in the 1960s and 70s our understanding of the manipulation of this structure type is still poor compared to perovskites.

The TTB structure, $A_2A'_4C_4B_2B'_8O_{30}$, is closely related to the perovskite structure, however, the presence of crystallographically non-equivalent A- and B-sites and an extra C-site provide extra degrees of freedom for manipulation of the structure offering huge compositional flexibility.[11] The TTB structure consists of a network of corner sharing $BO_6$ octahedra, figure 1 inset, which form perovskite (A) and pentagonal (A') channels which can be occupied by alkali, alkaline earth and rare earth cations. Smaller triangular (C) channels are also formed and whilst in many TTBs these sites are vacant they can be filled (or partially filled) by small low charged cations such as $Li^+$ (e.g. $K_6Li_4Nb_{10}O_{30}$[12]).

Much of the focus on recent TTB materials has been directed at $Ba_6Ti_2Nb_8O_{30}$[13] and its doped analogues as a result of both $Ti^{4+}$ and $Nb^{5+}$ being considered to be ferroelectrically active ions and the high Curie temperatures, $T_C$, (~500 K) observed in these materials. Isovalent replacement of the $Ba^{2+}$ cation with either $Sr^{2+}$ and/or $Ca^{2+}$ has been shown to result in changes in symmetry and thus shift the value of $T_C$.[14] Neurgaonkar *et al.* reported that $Sr_6Ti_2Nb_8O_{30}$ was orthorhombic in structure in comparison with $Ba_6Ti_2Nb_8O_{30}$ which is tetragonal[14] with a minimum in $T_C$ observed at ca. 70 mol % Sr substitution which they associated with a morphotropic phase boundary between the Ba-rich tetragonal phases and orthorhombic Sr-rich compositions. Furthermore, they demonstrated that the addition of $Ca^{2+}$ stabilised the orthorhombic distortion. The effects of incorporation of larger valence cations such as $Bi^{3+}$ and $RE^{3+}$ (RE = La, Pr, Nd, Sm, Eu, Gd or Dy) onto the A-site, through cooperative doping with $Ti^{4+}$ on the B-site, has also been reported.[15-18] It has been shown that whilst incorporation of $La^{3+}$ or $Bi^{3+}$ into the structure results in relaxor-type ferroelectric character incorporation of the smaller rare earths results in classic ferroelectric behaviour.[15, 16] More recently it has been suggested that the ferroelectric behaviour in these materials is dominated by the A-site cations and in particular by the ionic radius difference between the ions on the A1 and A2 sites.[19]

The incorporation of non-ferroelectrically active species, such as $Fe^{3+}$, $Ni^{2+}$ and $Mg^{2+}$, into the TTB framework has also been reported.[20-23] For rare earth substituted $Ba_6FeNb_9O_{30}$ studies have shown the size of the RE ion to have a large influence on the observed properties.[24, 25] However, in contrast little work has been undertaken in understanding the effects of B-site doping on the observed ferroelectric and dielectric properties. Replacement of $Ti^{4+}$ in ferroelectric $Ba_6Ti_2Nb_8O_{30}$ TTBs with $Sn^{4+}$, $Zr^{4+}$ or $Hf^{4+}$ has demonstrated that $T_C$ decreases with increasing cell volume ($M^{4+}$ ionic radius).[13, 14] In contrast, we recently showed that increasing the ionic radii of $M^{3+}$ ions in $Ba_6M^{3+}Nb_9O_{30}$ TTBs (M = $Ga^{3+}$, $Fe^{3+}$, $Sc^{3+}$, $In^{3+}$ of $Y^{3+}$) results in an increase in the polar stability.[26] These observations can more readily be described by tetragonality (c/a) such that irrespective of the valence of the metal ion in the TTB the transition temperature increases with increasing tetragonality (strain).[26] These observations marked steps towards quantitatively correlating the compositional-structure-property relations in TTBs so that a 'global' understanding of these materials similar to those in perovskites can be utilised in the design of new materials. This is an important step if novel materials of this structure type are to be fully exploited for future applications.

In this paper we report full structural and dielectric characterisation of $Ba_6M^{3+}Nb_9O_{30}$ ($M^{3+}$ = Ga, Sc or In) TTB materials. All materials exhibit dielectric permittivity and loss curves typical of relaxor behaviour. Characteristic temperature parameters for each sample were extracted from both dielectric and crystallographic data as a function of temperature: Vogel-Fulcher fitting of the maximum in dielectric constant to determine the Vogel-Fulcher temperature, $T_{VF}$; $T_{UDR}$ corresponding to absolute flattening of the dielectric loss peak in the frequency domain; and $T_{c/a}$ corresponding to the maximum crystallographic tetragonal strain. These temperatures coincided for each composition investigated and furthermore increased with increasing strain induced by an increase in the $M^{3+}$ cation size. These observations are consistent with slowing of dipoles on cooling and eventual "locking" of the B-cation displacements along the c-axis, i.e. dipole freezing.

**II Experimental**

Samples with the composition $Ba_6M^{3+}Nb_9O30$ (where $M^{3+}$ = Ga, Sc or In) were prepared as described previously.[26] Briefly, stoichiometric ratios of $BaCO_3$, $Nb_2O_5$, $Ga_2O_3$ $In_2O_3$ (all Aldrich, 99+ %) and $Sc_2O_3$ (Stanford Materials Corporation, 99.999 %) were ball milled and subjected to the following heating regime: 4 hours at 1000 °C, 10 hours at 1250 °C followed by ball milling and 6 hours at 1350 °C (1300 °C for $Ba_6GaNb_9O_{30}$). Phase formation was confirmed by powder x-ray diffraction collected over a range of 20 to 60 °2θ ≤ 60 in transmission mode using a STOE diffractometer with Cu $K\alpha_1$ radiation (40 kV and 30 mA, $\lambda$ = 1.5405 Å). Powder neutron diffraction measurements were conducted on the high resolution powder diffractometer (HRPD) at ISIS. Data were collected at temperatures between 10 and 450 K using a closed cycle refrigerator (CCR) with the materials loaded into aluminium cells fitted with vanadium windows. Pellets were prepared for electrical characterisation at 1350 °C (1300 °C for $Ba_6GaNb_9O_{30}$) and all exhibited > 90 % of theoretical density. Pt electrodes were applied using Pt paste (Gwent Electronic Materials Ltd.) and cured at 900 °C for approximately 20 minutes. Dielectric measurements were made using an Agilent 4294A impedance analyser over a frequency range of *ca.* 100 Hz – 5 MHz and a temperature range of approximately 50 – 340 K with cooling/heating rates of 2 $Kmin^{-1}$ and applied *ac* excitation of 500 mV.

## III Results

### A. General observations

Room temperature powder neutron diffraction (PND) indicated that all three compounds form the TTB phase with a small amount of minor perovskite related phase, $Ba_5Nb_4O_{15}$. All three were refined in the space group P4/mbm (No. 127) using the Rietveld method, figure 1 and table 1 (see EPAPS Document No. xxx for full crystallographic data). A systematic increase in lattice parameters and unit cell volume was observed with increasing ionic radius of the $M^{3+}$ B-site cation ($r_{Ga}^{3+} = 0.62 < r_{Sc}^{3+} = 0.745 < r_{In}^{3+} = 0.80$ Å for 6-fold coordination[27]), in good agreement with our previous x-ray diffraction study.[26] The amount of second phase was determined from dual phase refinements and was found to be approximately 10 % in $Ba_6GaNb_9O_{30}$ and 6 % in $Ba_6ScNb_9O_{30}$. No secondary phases were observed in the neutron data collected for $Ba_6InNb_9O_{30}$. Attempts to completely remove $Ba_5Nb_4O_{15}$ secondary phase by varying synthetic conditions was unsuccessful, however, the properties of this phase is well known[28]; it is an insulator with a temperature independent dielectric constant of *ca.* 40[28, 29] and so has negligible contribution to the observed electrical properties of the three compositions studied.

Dielectric spectroscopy data in the range 50 to 400 K showed that all three compounds exhibited characteristic relaxor behaviour with a strong frequency dependence of peaks in both dielectric constant and dielectric loss as a function of temperature. The dielectric curves were displaced to higher temperature with increasing average B-site ionic radii, as observed previously[26]. In order to further characterise this relaxor behaviour data were collected with higher temperature and frequency resolution than in our previous study. Dielectric data for the Sc-analogue, $Ba_6ScNb_9O_{30}$, are shown in figure 2.

**B. Analysis of dielectric data**

1) Vogel-Fulcher Analysis

The most commonly used method to evaluate the frequency response of the real part of dielectric permittivity in relaxors is the Vogel-Fulcher expression which was first adopted by Viehland[30] in the study of the archetypal relaxor ferroelectric, $PbMg_{1/3}Nb_{2/3}O_3$ (PMN), although there is some debate as to the validity of this approach, primarily due to problems with the goodness of fit and/or physically unreasonable fitted parameters, and this will be discussed in more detail in section IV, below. The Vogel-Fulcher (VF) model essential describes a temperature dependence of a spectrum of relaxation times and so probes the dynamics and population profile of the dipolar response(s) as a function of temperature. The VF equation (1) is simply a modified Arrhenius expression which includes an increasing degree of interaction between random local relaxation processes, in this case of the dipolar response:

$$f = f_0 \exp\left( -\frac{E_a}{k(T_m - T_{VF})} \right) \tag{1}$$

where $f$ is the frequency of the perturbation (applied *ac* field frequency, Hz); $f_0$ is the fundamental attempt or limiting response frequency of the dipoles (Hz); $E_a$ is the activation energy of local polarization (J); $T_m$ is the temperature (K) of maximum dielectric constant at frequency, $f$; $T_{VF}$, the characteristic Vogel-Fulcher temperature (often described as the static freezing temperature[30] (K)); and $k$ is Boltzmann's constant ($1.381 \times 10^{-23}$ $JK^{-1}$).

Fits of $T_m$ data obtained from dielectric constant curves (e.g. figure 2a) to the VF expression (equation 1) are shown in figure 3 for all three compositions. In each case the extracted parameters are all physically sensible, as shown in table 2: the fundamental attempt frequency, $f_0$, is of the order of $10^{12}$ Hz (0.67, 0.09 and 0.40 THz for Ga, Sc and In analogues, respectively); and both the activation energy, $E_a$, and Vogel-Fulcher temperature, $T_{VF}$, systematically increase with increasing B-cation size, in agreement with the less quantitative observation that the dielectric curves are displaced to higher temperature with increasing cation size.

2) Fitting of dielectric loss

Vogel-Fulcher analysis often results in unreliable and unphysical values for the fitting parameters[19, 31-33] primarily due to the sensitivity of fitting to the curvature of $T_m(f)$ data (usually obtained over a limited frequency range, e.g. $10$-$10^6$ Hz) and the subsequent extrapolation which is often over several orders of magnitude; our experience has also found that sample quality is also of critical, particularly for polycrystalline ceramics where microstructural factors such as grain size and sample density can have a dramatic affect on the values of $T_m$ (this will be discussed in a subsequent manuscript). In order to verify the parameters obtained by VF fitting of the real part of dielectric permittivity the imaginary component (dielectric loss, $\varepsilon''$) was also analysed. The dielectric data in figure 2 a-b are plotted in the frequency domain in figure 2 c-d, and as a complex plane, Cole-Cole[34], plot, figure 2e. The data are consistent with that observed in other relaxors, including relaxor ferroelectrics[35-38]. The theory of relaxors predicts that at the static freezing temperature the dielectric loss peak associated with the dipolar response should become infinitely broad (flat) in the frequency domain. The most simple model to analyse dielectric relaxation is the simple Debye response which contains a single relaxation time; this model however results in a symmetric loss peak with a unity relationship with frequency, $\varepsilon'' \propto f^n$ where $n$ = -1 and +1 below and above the relaxation frequency, $f_p$, respectively. In reality systems show more dispersive behaviour and a Gaussian distribution of relaxation times (DRT) or Cole-Cole expression[34] is often introduced to fit broadened dielectric loss peaks, however these remained symmetric about the peak relaxation frequency. The dielectric loss data as a function of frequency, figure 2d, and Cole-Cole plots, figure 2e, indicate a high degree of asymmetry so these analytical methods are not suitable. There are a number of models to allow for asymmetric loss peaks such as the Cole-Davidson[39] or Havriliak-Negami[40] expressions, however for simplicity we have used the empirical two exponent model of Jonscher's universal dielectric response (UDR)[41] in order to fit this $\varepsilon''(f)$ data. This approach describes the frequency dependence of dielectric loss by the expression:

$$\varepsilon'' \propto \frac{1}{(f/f_p)^{-m} + (f/f_p)^{1-n}} \tag{2}$$

where $f$ is the *ac* field frequency ($=\omega/2\pi$); $f_p$ is the relaxation (peak) frequency; and $-m$ and $(1-n)$ are the frequency exponents of $\varepsilon''(f)$ below and above $f_p$, respectively, with the conditions $0 \leq m, n \leq 1$. This empirical model allows a Debye response for $m = 1$, $n = 0$, and a DRT (Cole-Cole) response for $(m = (1-n)) < 1$. Furthermore, Jonscher's model also predicts that the temperature dependence of the frequency of maximum dielectric loss, $f_p$, (or dielectric relaxation frequency which is related to the relaxation time, $\tau = \omega_p^{-1} = (2\pi f_p)^{-1}$ where $\omega_p$ is the angular frequency of the peak maxima) can be described by the Arrhenius expression:

$$f_p = f_0 \exp\left(-\frac{E_a}{kT}\right) \tag{3}$$

where $f_0$ is limiting dipole response frequency and $E_a$ is the activation energy. Thus by fitting dielectric loss data to equations (2) and (3), it is possible to determine some characteristic temperature, $T_{UDR}$, corresponding to slowing of the longest mean relaxation time by extrapolation to zero of the gradient, $m$ ($\varepsilon'' \propto f^{-m}$ for $f < f_p$), and $f_0$ and $E_a$ from the temperature dependence of $f_p$, respectively.

Fits to the UDR model of dielectric loss data at selected temperatures for $Ba_6ScNb_9O_{30}$ are shown in figure 4a. The temperature dependence of the gradient, m, on the low frequency side of the $\varepsilon''(f)$ peaks is shown in figure 4b and indicates a steady decrease in $m$ with decreasing temperature. Absolute flattening of the loss peak (i.e. m $\rightarrow$ 0) is predicted to occur at the "dipole freezing temperature"[42, 43]; extrapolation of the m data zero gives $T_{UDR} \approx 150$ K, in comparison to $T_{VF}$ of 153 K as determined from VF fitting of dielectric constant data. Arrhenius fitting of the temperature dependence of the peak frequency, $f_p$, is shown in figure 4c, and gives values of $3.20 \times 10^{11}$ Hz and 0.09 eV for $f_0$ and $E_a$, respectively. These are presented in table 2 alongside those determined from VF analysis. This independent treatment of both parts of the complex dielectric permittivity data give characteristic temperature ($T_{\mathrm{VF}}$ or $T_{UDR}$), limiting frequency, $f_0$, and activation energy, $E$, parameters; these are summarised in table 2. The rationalisation of application of both Vogel-Fulcher and Arrhenius behaviour is discussed in section IV.

Due to the limited frequency and temperature range available and also noise in the data it was not possible to accurately determine $f_p(T)$ data for Ga and In analogues in order to determine $f_0$ and $E_a$ using equation 3, however it was possible to determine $m$ from the gradient of the $\varepsilon''(f)$ curves over a wide temperature range. Both samples showed similar behaviour to the Sc-compound with $m$ decreasing with decreasing temperature, figure 4d. Extrapolation of $m(T)$ curves to $m$ = 0, gave $T_{UDR}$ = 58 and 183 K, for $Ba_6GaNb_9O_{30}$ and $Ba_6InNb_9O_{30}$, respectively, compared to $T_{VF}$ values of 56 and 158 K from VF fitting, figures 3a and c. It should be noted there is some uncertainty in the value of $T_{UDR}$ for the In-analogue; this is discussed in section IV.

**C. Temperature dependence of crystal structure**

Variable temperature PND in the range 450 to 8 K was used to investigate any changes in crystal structure. While some relaxor ferroelectrics such as PZN exhibit a macroscopic symmetry change[44] others such as PMN display only local, randomly aligned (under zero field conditions) distortions associated with local non-centrosymmetric polar nano-regions which result in diffuse scattering around the Bragg peaks[45, 46]. There was no evidence of any long range symmetry lowering at any temperature in any of the three TTB samples studied here; all data were therefore refined in the tetragonal, centrosymmetric space group *P4/mbm*. The evolution of individual lattice parameters is shown in figure 5. Refinement of site occupancies of $M^{3+}$ and $Nb^{5+}$ suggested a completely disordered system and this was supported by a significantly poorer fit to models where $M^{3+}$ cations were constrained to either B1 or B2 sites. In all cases the materials exhibit a linear contraction in the *ab* plane with linear coefficients, $\alpha$, typical of oxides ($\alpha$ = 7.92, 8.15 and 8.42 $\times$ $10^{-6}$ $K^{-1}$ for Ga, Sc and In analogues, respectively) down to *ca.* 100 K below which the contraction slows. As expected from the relative magnitude of the longer *a, b* axes compared to the *c* axis (*ca.* 12.6 Å *vs.* 4 Å), this behaviour dominates the volume behaviour, figures 5 d-f. The *c*-axis data, however, exhibits a dramatic deviation from the expected linear behaviour, figures 5 a-c; in each case the rate of contraction is much lower from room temperature to around 150, 200 and 250 K for Ga, Sc and In samples, before dramatically increasing, as indicated by the strong inflexion in the data. In the approximately linear region below the inflexion $\alpha$ = 5.94, 5.75 and 6.18 $\times$ $10^{-6}$ $K^{-1}$ for Ga, Sc and In analogues, respectively. This highly anisotropic contraction is a clear indication of some subtle structural change. It is important to note that this behaviour merely reflects the average change in shape and size of the (tetragonal) unit cell and in no way reflects the local position of the atoms; this is discussed in more detail below. In our previous study we noted that the degree of anisotropy and therefore strain (quantified by the tetragonality, i.e. ratio of *c/a*) showed an approximately linear relationship with $T_m$ or $T_C$ for B-site substituted relaxor or ferroelectric TTBs[26]. The crystallographic data in figures 5 a-c was re-plotted as tetragonality (*c/a*), figure 6. All samples undergo a point of maximum tetragonality (maximum anisotropy) as a function of temperature; in each case this characteristic temperature, $T_{c/a}$, correlates with the temperatures, $T_{VF}$ and $T_{UDR}$, as determined from electrical analysis. A summary of the three extracted temperature parameterss, $T_{VF}$, $T_{UDR}$ and $T_{c/a}$, obtained from the three methods is presented in table 3.

## IV Discussion

*Systematic correlation with B-cation size*

The relaxor properties of samples were initially probed using the most common approach of fitting of the real part of dielectric permittivity to the Vogel-Fulcher model. This allowed extraction of three variables, namely the Vogel-Fulcher temperature, $T_{VF}$, activation energy, $E_a$, and fundamental response frequency, $f_0$. Where possible $E_a$ and $f_0$ parameters and characteristic temperature, $T_{UDR}$, were also determined by fitting of the dielectric loss to Jonscher's empirical two-exponent UDR model[41]. Parameters obtained from analysis of both real and imaginary parts of the complex dielectric function give good agreement, table 3. The largest discrepancy is the value obtained for $T_{UDR}$ of the In analogue as determined from the extrapolation of *m* to zero, figure 4d; this method gave a significantly higher value than the temperature extracted from VF fitting and the crystallographic data. The extrapolation of the *m* data for this sample is from well above the expected intercept and assumes linear behaviour. Similar *m* data for the Sc sample, for which it was possible to obtain *m* values over a wider temperature range, showed clear non-linear behaviour, figure 4b; if this is also the case for the In sample, then the extrapolation used here is a significant over-estimate of $T_{UDR}$. Unfortunately it was not possible to obtain *m* at lower temperatures in this instance for the In-analogue.

Crystallographic data showed a clear structural anomaly, manifested as a maximum in anisotropy, and hence tetragonal strain at temperatures corresponding those temperature parameters determined by two different analysis methods of dielectric data. The structural origin of this behaviour is discussed below, but what is clear is that all three samples show consistent and systematic behaviour: with increasing B-cation radius the dielectric curves are displaced to higher temperature. More in-depth analysis shows a systemic increase in the parameters $E_a$ and, more pertinently, characteristic temperatures, $T_{VF}$, $T_{UDR}$ and $T_{c/a}$ as determined from 3 different methods. The latter show an almost linear trend as a function of $r_B$, figure 7, as we suggested previously[26]. The overall behaviour is entirely consistent with dipoles generated by an off-centring of cations which decreases in magnitude with decreasing thermal energy (temperature). For the electrical response the decreasing mobility of the atomic displacements is manifested as a slowing down of the average relaxation frequency as noted in the $\varepsilon''(f,T)$ response. At $T_{c/a}$ the atoms no longer have sufficient energy to displace through the centrosymmetric position and so are locked within an energy minima in a certain crystallographic axis resulting in stiffening of the lattice and a resistance to thermal contraction. This atomistic explanation is examined in more detail in terms of the crystallographic data below. If the characteristic temperature parameters, $T_{VF}$, $T_{UDR}$ and $T_{c/a}$ do indeed correspond to some dipole freezing temperature, $T_f$, then it denotes the thermal stability of the dipoles, and is clearly related to the local crystal structure.

*Phyiscal manifestation of dipole freezing*

There is some considerable debate as to the physical meaningfulness of the parameter $T_{VF}$ (often refereed to as a dipole freezing temperature)[42, 47], both because in some cases the parameters obtained for relaxor ferroelectrics and dielectrics are physically unrealistic[31, 32] and also in many polymers and glasses, for example, $T_{VF}$ as determined from VF fitting corresponds to some seemingly arbitrary temperature below the glass transition temperature, $T_g$, determined by thermal analysis. One would naturally assume in the latter, glassy systems that $T_g$ and $T_{VF}$ would coincide as the $T_g$ describes the temperature at which "molecular" or segmental motion ceases (freezes).

It is clear in our materials, however, that $T_{VF}$ as determined from the macroscopic, but dynamic, dielectric response corresponds to a physical process which is manifested in the macroscopic, time- and spatially-averaged crystal structure determined by diffraction. This indicates that $T_{VF}$ relates to a physical process. The crystallography shows a clear and unambiguous relationship between maximal tetragonal strain and $T_{VF}$; this behaviour is entirely consistent with the freezing of dipoles which originate from atomic displacements.

In our analysis there remains the question as to why the dielectric constant data are best described by the Vogel-Fulcher expression (a modified Arrhenius with some finite thermal activation temperature, $T_{VF}$) while the dielectric loss displays simple thermally activated (Arrhenius) behaviour (although it should be noted that it was only possible to obtain $f_p$ values over a narrow temperature range, figure 4c). Such behaviour was previously observed in the related TTB $Sr_2LaTi_2Nb_3O_{15}$[36], however the authors clearly identified two contributing relaxations in the dielectric loss which they attributed to a breathing and flipping mechanism of PNRs. We do not see such a double contribution in our samples and so there are two possibilities for the observed behaviour: (i) there are two separate relaxation processes, one dominates the dielectric polarisability and the other the loss, but which are not resolvable; second that a single mechanism gives rise to the behaviour observed in both the real and imaginary parts. The first scenario may be unlikely given that the $T_{VF}$ and $T_{UDR}$ temperatures linked to two separate responses are coincident in each sample yet vary significantly across the compositional series. The second scenario can be rationalised based on the description at atomic displacements outlined in the previous section. The maxima in dielectric constant represent contributions of displacements through the centrosymmetric position (z=0.5 in the unit cell), i.e. flipping, and $T_{VF}$ represents the temperature at which they no longer have sufficient energy (thermal or field-driven) to do so; they become locked in the *c*-axis but are still prone to small perturbation both along the *c*-axis (but not through the point of inversion symmetry), i.e. breathing, and in the *ab* plane. This behaviour is akin to a flipping and breathing type model but for non- or weakly-interacting dipoles rather than collective behaviour within PNRs for which we have no evidence (see discussion below). This accounts for both a freezing type behaviour for polarisability (and hence Vogel-Fulcher) behaviour and finite but slowing relaxation of the dielectric loss at all temperatures.

*Structural origin of the dipolar response*

Many relaxor dielectrics differ from classical ferroelectrics in that they undergo no macroscopic change in symmetry even if there is local non-centrosymmetry as in the polar nano region (PNR) model, e.g. in PMN. Nevertheless, even in PMN there are some structural correlations associated with the nucleation and growth of such PNRs. For example the local rhombohedral distortion in the PNRs with polarization developed along the <111> axis results in weak and diffuse Bragg scattering.[45, 46] It also results in a resistance to linear thermal contraction as a result of the strain induced by the local dipoles. In zero-field-cooled (ZFC) PMN the strain induced during random nucleation of PNRs in the 8-fold <111> directions on cooling from the high temperature cubic polymorph spatially average resulting in an isotropic relative expansion[45]. While there is no evidence of any diffuse Bragg peaks, a similar non-linear volume contraction at low temperature is observed. However, this only occurs below 100 K (several hundred K lower than in PMN which shows a deviation starting at the Burns temperature which is associated with PNR nucleation, and which also results in deviation from Curie-Weiss behaviour and a change in refractive index)[48-51] and appears to occur at a similar temperature in all materials independent of the $M^{3+}$ ion and therefore most likely denotes the approaching limit for bond contraction. The crystallographic data also show, however, a clear structural anisotropy at $T_{c/a}$ which is dominated by the non-linear *c*-axis response. This clearly suggests that the dipoles (displacements) are (predominantly) active in the *c*-axis. It worthy to note that it is simply as a result of the highly anisotropic nature of the TTB structure that this distinction is clear; the majority of diffraction studies of relaxor dielectrics are perovskite based which have pseudo-cubic local distortions which are spatially averaged and so make identification of the displacement axis difficult using long range, structure averaging diffraction techniques. In order to further investigate any displacive origin of the dipoles the diffraction data were examined in more detail.

Initial refinements were carried out in the centrosymmetric, P4/mbm, space group with atomic displacement parameters refined isotropically ($U_{iso}$); these criteria confine the B-cations in the *c*-axis (to z = 1/2 in the unit cell) but allow displacement from their ideal positions in the *ab* plane (full crystallographic models may be found in EPAPS Document No. xxx). The refinements indicate that the $U_{iso}$ values for the B1 site and the oxygen ions that link the $MO_6$ octahedra in the c plane (O1 and O3) appear higher than those determined for the other ions within the unit cell. The data were refined again (in the P4/mbm space group) but with anisotropic atomic displacement parameters, $U_{aniso}$, for the two B-site positions. All refinements showed improved goodness of fit factors as a result of the extra degrees of freedom associated with $U_{aniso}$ (see EPAPS Document No. xxx for the full crystallographic refinements). The results clearly show a large anisotropy for the B1 cation with much larger displacements in the *c*-axis as indicated by $U_{33}$. Table 4 gives the goodness of fit factors and $U_{aniso}$ for the Nb1 and Nb2 positions for the $Ba_6GaNb_9O_{30}$, $Ba_6ScNb_9O_{30}$ and $Ba_6InNb_9O_{30}$ at 100 K (all other data is given in EPAPS Document No. xxx). This strongly suggests that the dipoles originate from displacement of B1 cations in the *c*-axis. In an attempt to estimate more closely the magnitude of these displacements, refinements were carried out in the non-centrosymmetric space group P4bm; the mirror plane perpendicular to the c-axis at $z = 0.5$ is removed allowing off-centring of the B-cations. These refinements were conducted fixing the A-site cations at the origin and in the *z* direction for the A1 and A2 sites, respectively, to allow the atomic displacements of the B-site cations to be quantified. There is very little movement of the $Ba^{2+}$ ion on the A2-site in either the *x* or *y* directions as a result of fixing these crystallographic sites. Similarly, very little displacement is observed for the $O^{2-}$ ions (particularly in the *z* direction) whilst there is a very clear displacement of the B-site cations.

The B-cation atomic displacements are shown visually in figure 8 at a temperature of 100 K. (In *P4/mbm* symmetry the B-cations ions are constrained on the solid line in the centre). Close inspection of the $MO_6$ octahedral units (figure 8) suggests that whilst the ions in B2-site are displaced predominantly in the *c*-direction they are also (more weakly) displaced in the *ab* plane. This suggests that rather than being significantly linked to dipolar behaviour this displacement arises as a result of structural distortion and a rotation of the $MO_6$ (B2) octahedra (see EPAPS Document No. xxx). On the other hand the B1 octahedral site shows a clear displacement in the *c*-direction consistent with a dipole moment. These results would suggest that the B1 site dominates the dipolar response in these materials. Unsurprisingly (as suggested by the $T_{c/a}$ values), the magnitude of displacement increases with increasing $M^{3+}$ ionic radii confirming the relationship between crystallographic strain and polar stability[26].

*Nature of the relaxor behaviour*

The precise nature of the relaxor behaviour in these materials is not currently known. There are two broad categories of relaxor depending on the degree and length scale of correlation between the dipoles: canonical relaxor dielectrics, which are essential dipole glasses with weak dipole-dipole interactions and little or no dipole ordering; and relaxor ferroelectrics with short range polar ordering resulting in ferroelectrically active nano-domains (commonly referred to as polar nano-regions, PNRs). The data reported here clearly indicate that this family of materials are essentially uniaxial relaxors with the dipolar response dominant in the short *c*-axis.

Recently, the random-field Ising model (RFIM) similar to that used to explain spin-glass systems has been used to describe uniaxial relaxor character in $Sr_{0.61}Ba_{0.39}Nb_2O_6$ (SBN) tetragonal tungsten bronzes[52]. In SBN the relaxor ferroelectric character arises as a result of the formation of metastable PNRs which are separated by paraelectric interfaces. The formation of PNR occurs as a result of A-site disorder driven not only by the size differences between $Ba^{2+}$ and $Sr^{2+}$ but also by the random distribution of vacant sites[52]. The formation/random distribution of these vacancies are suggested to create random electric fields which favour the alignment of polarised clusters. Our materials incorporate neither A-site disorder nor vacancies. Furthermore, PND data suggests that the dipolar response in these materials is predominantly linked to the B1-site making these materials chemically dilute (the ratio of B1- to B2-sites is 1:4 in the TTB unit cell). This would suggest that the RFIM does not accurately describe the origin of the relaxor behaviour in $Ba_6M^{3+}Nb_9O_{30}$ materials. In these TTBs it can be considered that we have a dilute dipolar system arising from ‘polar’ chains of B1-ions spatially separated within an effectively non-polar/weakly polar matrix formed by the crystallographic distribution of the B2-sites. Whilst the Ising model describes uniaxial relaxor behaviour it is based on domain formation (as described above) and thus cannot describe the behaviour seen here since the formation of large polar domains is restricted by the crystallography. In an attempt to investigate any relaxor ferroelectric properties in these materials polarisation-field (P-E) measurements were carried out on the $Ba_6InNb_9O_{30}$ samples at 288-298 K (i.e. well within the range of $T_m$ values) in the frequency range 1 Hz – 1 kHz and at applied fields of up to 50 kV/cm; the response was consistent with a (slightly lossy) linear dielectric. In relaxor ferroelectrics such as PMN even at $T_m$, and above, the residual PNRs result in either a slim loop or non-linear behaviour[53]. We observe neither, although the coercive field in these materials may be as high as 100 kV/cm[54] which we were unable to sustain and so these results are far from conclusive. In addition, howevere field cooled (1 kV/cm) followed by zero field heating dielectric measurements from ambient to 50 K in the Sc analogue showed no change in the dielectric response. The absence of PNRs is also supported by the Curie-Weiss plots (EPAPS Document No. xxx) which show no deviation associated with PNR nucleation at some Burn’s temperature, $T_{d.}$ While certainly not conclusive, the absence of any hysteresis (P-E) loop or non-linear behaviour combined with the lack of variation of dielectric response on field cooling and also the excellent Curie-Weiss behaviour collectively suggest an absence of PNRs i.e. the behaviour is not dominated by nucleation and growth, but more of a glassy freezing character; $T_f$ could therefore represent the dipole freezing temperature denoting a transition from the ergodic into a nonergodic state characterized by divergence of the longest relaxation time[55]. This clearly favours a canonical relaxor dielectric (dipole glass) model for these materials, but this requires further investigation. Lastly, we have not investigated the possibility of an incommensurate crystal structure in our materials. Such periodic modulations result in a distribution of dipole environments and hence relaxor behaviour[56, 57].

**V Conclusions**

The relaxor properties of a previously reported[26] family of dielectrics, $Ba_6M^{3+}Nb_9O_{30}$ with $M^{3+} = Ga^{3+}$, $Sc^{3+}$ and $In^{3+}$, are presented in detail. The dielectric response was analysed using Vogel-Fulcher fitting of the frequency dependence of the maxima in real part of the dielectric permittivity to obtain the characteristic parameters denoted fundamental dipole response frequency, $f_0$, activation energy, $E_a$, and Vogel-Fulcher temperature, $T_{VF}$. Where possible $f_0$, $E_a$ and characteristic temperature, $T_{UDR}$, parameters were also determined by independent fitting of the dielectric loss to Jonscher's empirical two exponent model. In all cases the values obtained were in good agreement and were also physically sensible. Given the particular sensitivity of the VF methodology and its propensity to return non-physical values, the latter method may provide a useful verification or even alternative approach for characterisation of relaxor behaviour. Overall, the analysis of dielectric data showed a systematic (almost linear) increase in the dipolar response of these materials with increasing $M^{3+}$ cation size.

Crystallographic studies using variable temperature power neutron diffraction were used to probe the structural response and showed that all three materials undergo non-linear thermal contraction in the short *c*-axis resulting in a maxima in tetragonality at a characteristic temperature, $T_{c/a}$, correlating also to $T_{VF}$ and $T_{UDR}$ determined from dielectric data. Rietveld refinements show that the dipolar response is dominated by local, non-cooperative displacement of B1 cations in the *c*-axis. The magnitude of displacement increases with average B-cation size and therefore crystal anisotropy and is entirely consistent with the observed dielectric properties which suggest a linear increase in $T_{c/a}$, $T_{VF}$ and $T_{UDR}$ with B-cation size. The collective behaviour suggests that the characteristic temperatures $T_{c/a}$, $T_{VF}$ and $T_{UDR}$ determined form the various analysis is consistent with the slowing and eventual freezing of a dipolar response in these materials at a temperature $T_f$.

The crystallography unambiguously shows that in this family of materials $T_f$ relates to a real physical process manifested as maximal crystallographic strain in the active dipolar axis and so is entirely consistent with dipole freezing. In these materials, therefore, $T_f$ may be used as a direct metric of thermal stability of the dipoles.

Collectively, the data also strongly suggest that these materials are not relaxor ferroelectrics, but are better described by a "dipole glass"-type model and may therefore be better classified as relaxor dielectrics.

**Acknowledgements**

The authors would like to thank the following funding organisations for support: the Royal Society for the provision of a research fellowship (for FDM); the EPSRC (DCA and AR); STFC for access to neutron facilities; Roberto Rocca Education Program and INFLPR Bucharest (AR). We would also like to thank Prof Phil Lightfoot for discussions regarding crystallographic data.

**Table 1.** Unit cell dimensions and goodness of fit parameters for $Ba_6M^{3+}Nb_9O_{30}$ ($M^{3+}$ = $Ga^{3+}$, $Sc^{3+}$, $In^{3+}$) refined from PND data at 300 K in space group *P4/mbm*.

| | $a$ (Å) | $c$ (Å) | $V$ (Å$^3$) | $R_p$ (%) | $wR_p$ (%) | $\chi^2$ |
|---|---|---|---|---|---|---|
| $Ba_6GaNb_9O_{30}$ | 12.5723(3) | 3.98181(2) | 629.384(4) | 6.04 | 6.19 | 13.46 |
| $Ba_6ScNb_9O_{30}$ | 12.63011(2) | 4.00746(1) | 639.268(3) | 4.35 | 4.77 | 8.542 |
| $Ba_6InNb_9O_{30}$ | 12.64713(2) | 4.01688(1) | 642.500(3) | 4.31 | 4.81 | 7.248 |

**Table 2.** Vogel Fulcher and goodness of fit parameters for $Ba_6M^{3+}Nb_9O_{30}$ ($M^{3+}$ = $Ga^{3+}$, $Sc^{3+}$, $In^{3+}$) determined from data fitting in figure 3.

| | $T_{VF}$ (K) | $f_0$ (Hz) | $E_a$ (eV) | *R.M.S.D.*† | $\chi^2$ |
|---|---|---|---|---|---|
| $Ba_6GaNb_9O_{30}$ | 56.3 | $6.70 \times 10^{11}$ | 0.0768 | 0.0598 | 0.0608 |
| $Ba_6ScNb_9O_{30}$ | 152.9 | $9.10 \times 10^{10}$ | 0.1052 | 0.1158 | 0.2549 |
| $Ba_6InNb_9O_{30}$ | 158.3 | $4.05 \times 10^{11}$ | 0.1977 | 0.0532 | 0.0455 |

† Root mean standard deviation.

**Table 3.** Summary of *temperature parameters* determined from fitting of dielectric data to Vogel-Fulcher ($T_{VF}$) and universal dielectric response ($T_{UDR}$) models and from maximum tetragonality ($T_{c/a}$) determined from crystallographic data.

| | $T_{c/a}$ (K) | $T_{VF}$ (K) | $T_{UDR}$ (K) |
|---|---|---|---|
| $Ba_6GaNb_9O_{30}$ | 75±15 | 56.3 | 58±1.5 |
| $Ba_6ScNb_9O_{30}$ | 145 | 152.9 | 150±2.5 |
| $Ba_6InNb_9O_{30}$ | 158 | 158.3 | 183±3.5 |

**Table 4.** Goodness of fit parameters and $U_{aniso}$ for the Nb1 and Nb2 sites refined in the space group P4/mbm showing the large displacement in the $U_{33}$ parameter (particularly on the B1 site) consistent with the anisotropy observed in the *c*-axis.

| Parameters | $Ba_6GaNb_9O_{30}$ | $Ba_6ScNb_9O_{30}$ | $Ba_6ScNb_9O_{30}$ |
|---|---|---|---|
| $\chi^2$ | 8.526 | 10.49 | 8.981 |
| wRp (%) | 4.89 | 5.27 | 5.36 |
| Rp (%) | 4.63 | 4.85 | 4.72 |
| a (Å) | 12.55351(3) | 12.61249(3) | 12.62929(3) |
| c (Å) | 3.97861(2) | 4.00406(1) | 4.01288(2) |
| Cell Volume (Å$^3$) | 626.992(3) | 636.945(3) | 640.051(3) |
| Nb1/$M^{3+}$1, (0,½, ½) | | | |
| Nb1/ $M^{3+}$1 U(aniso) × 100 Å$^2$ | | | |
| $U_{11}$ | 0.43(5) | 0.40(5) | 0.02(5) |
| $U_{22}$ | 0.43(5) | 0.40(5) | 0.02(5) |
| $U_{33}$ | 2.1(1) | 3.18(9) | 2.8(1) |
| $U_{12}$ | -0.07(7) | -0.03(6) | 0.06(7) |
| $U_{13}$ | 0.0 | 0.0 | 0.0 |
| $U_{23}$ | 0.0 | 0.0 | 0.0 |
| Nb2/ $M^{3+}$2 (x,y, ½) | 0.07472(7) | 0.07556(6) | 0.07605(7) |
| | 0.21450(6) | 0.21421(5) | 0.21412(7) |
| Nb2/ $M^{3+}$2 U(aniso) × 100 Å$^2$ | | | |
| $U_{11}$ | 0.64(4) | 0.43(4) | 0.56(5) |
| $U_{22}$ | 0.13(4) | 0.08(4) | 0.22(4) |
| $U_{33}$ | 1.00(4) | 0.92(3) | 1.04(4) |
| $U_{12}$ | 0.11(4) | 0.12(3) | 0.29(4) |
| $U_{13}$ | 0.0 | 0.0 | 0.0 |
| $U_{23}$ | 0.0 | 0.0 | 0.0 |

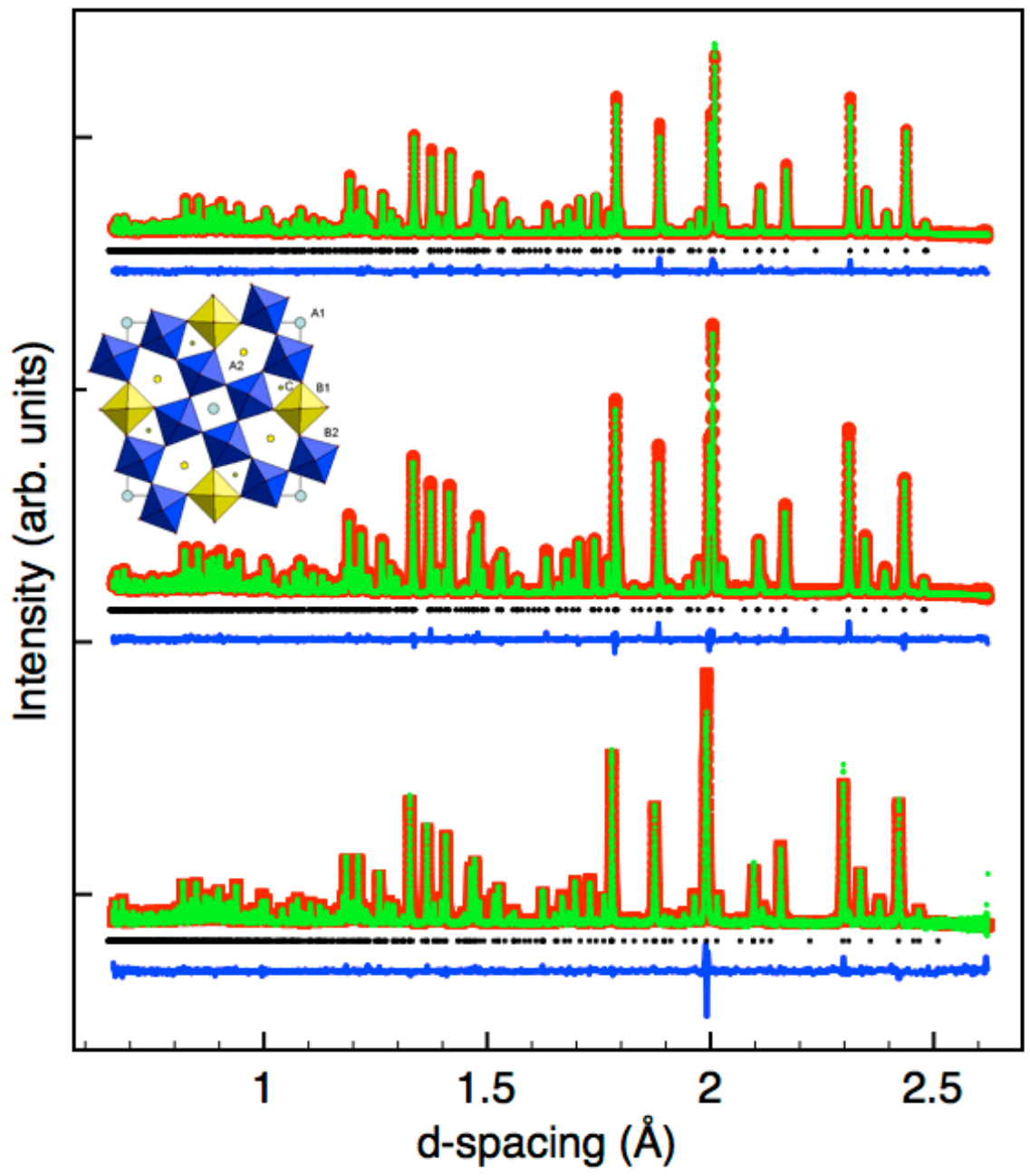

**Figure 1** (Colour online) Rietveld refinements in centrosymmetric tetragonal space group P4/mbm of room temperature powder neutron data for $Ba_6GaNb_9O_{30}$ (bottom), $Ba_6ScNb_9O_{30}$ (middle), and $Ba_6InNb_9O_{30}$ (top). (Inset - TTB structure with B1 and B2 octahedra shown in yellow and blue, respectively)

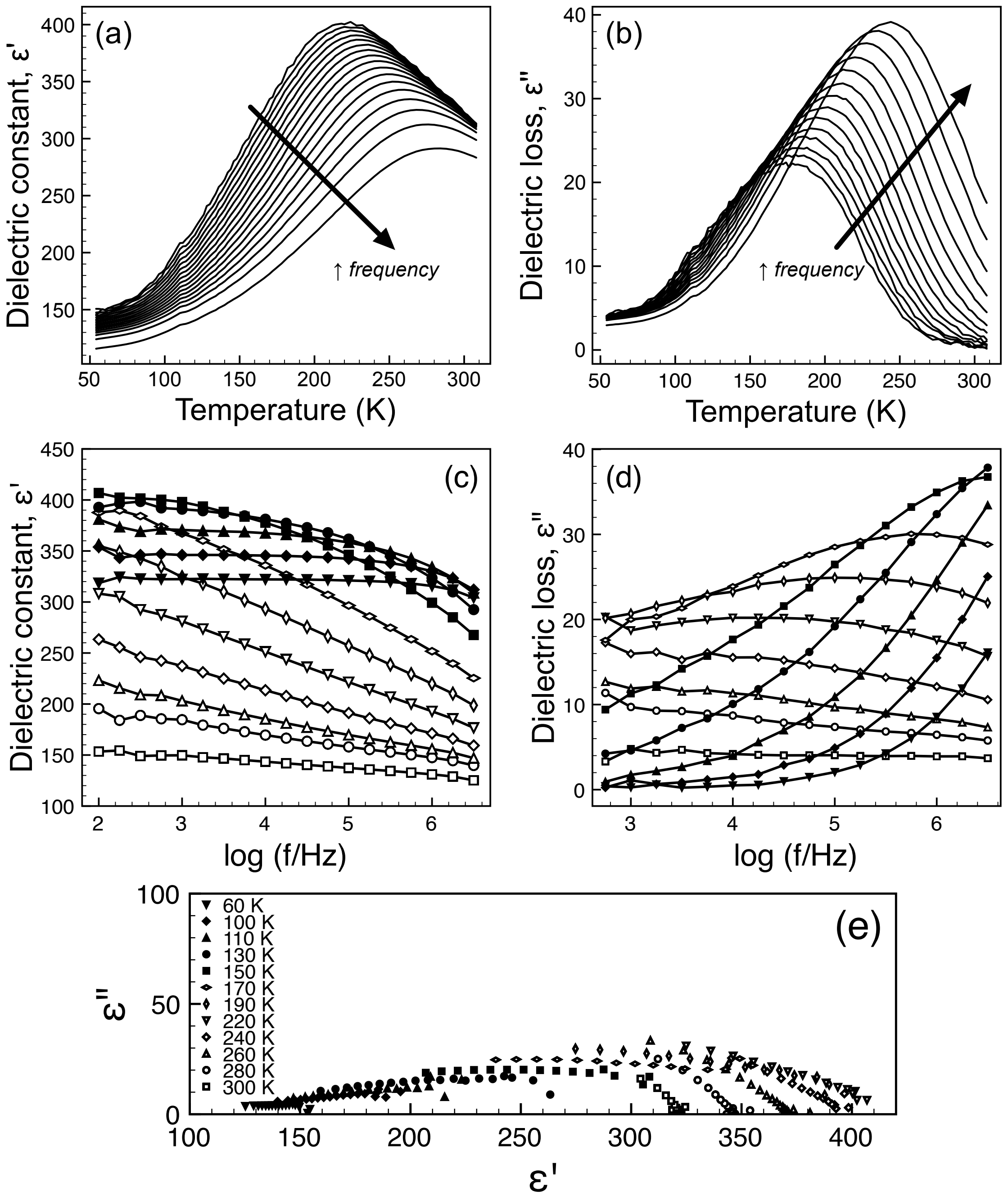

**Figure 2** Dielectric constant (a) and loss (b) as a function of frequency and temperature for $Ba_6ScNb_9O_{30}$. Real (c) and imaginary (d) components of dielectric permittivity as a function of frequency at various temperatures, and also as a Cole-Cole plot (e). The data key in part (e) also applies to parts (c) and (d).

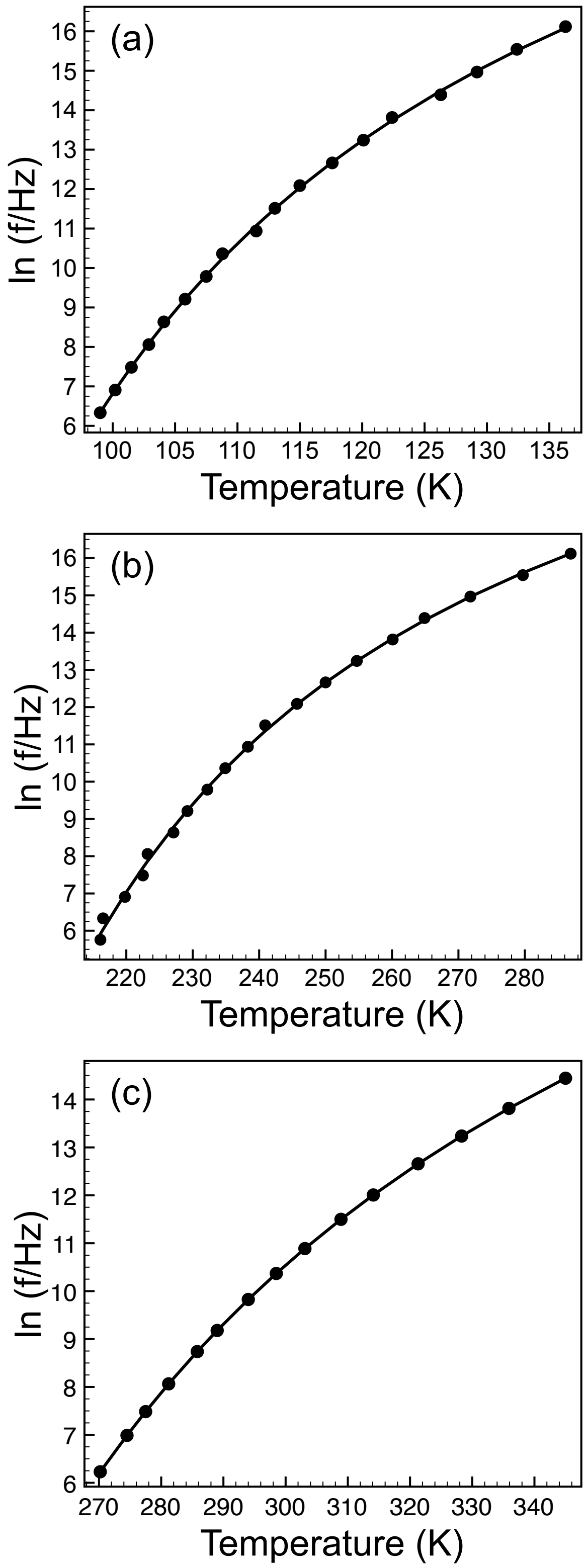

**Figure 3** Vogel-Fulcher fits of $T_m(f)$ extracted from dielectric constant data for $Ba_6GaNb_9O_{30}$ (a), $Ba_6ScNb_9O_{30}$ (b), and $Ba_6InNb_9O_{30}$ (c).

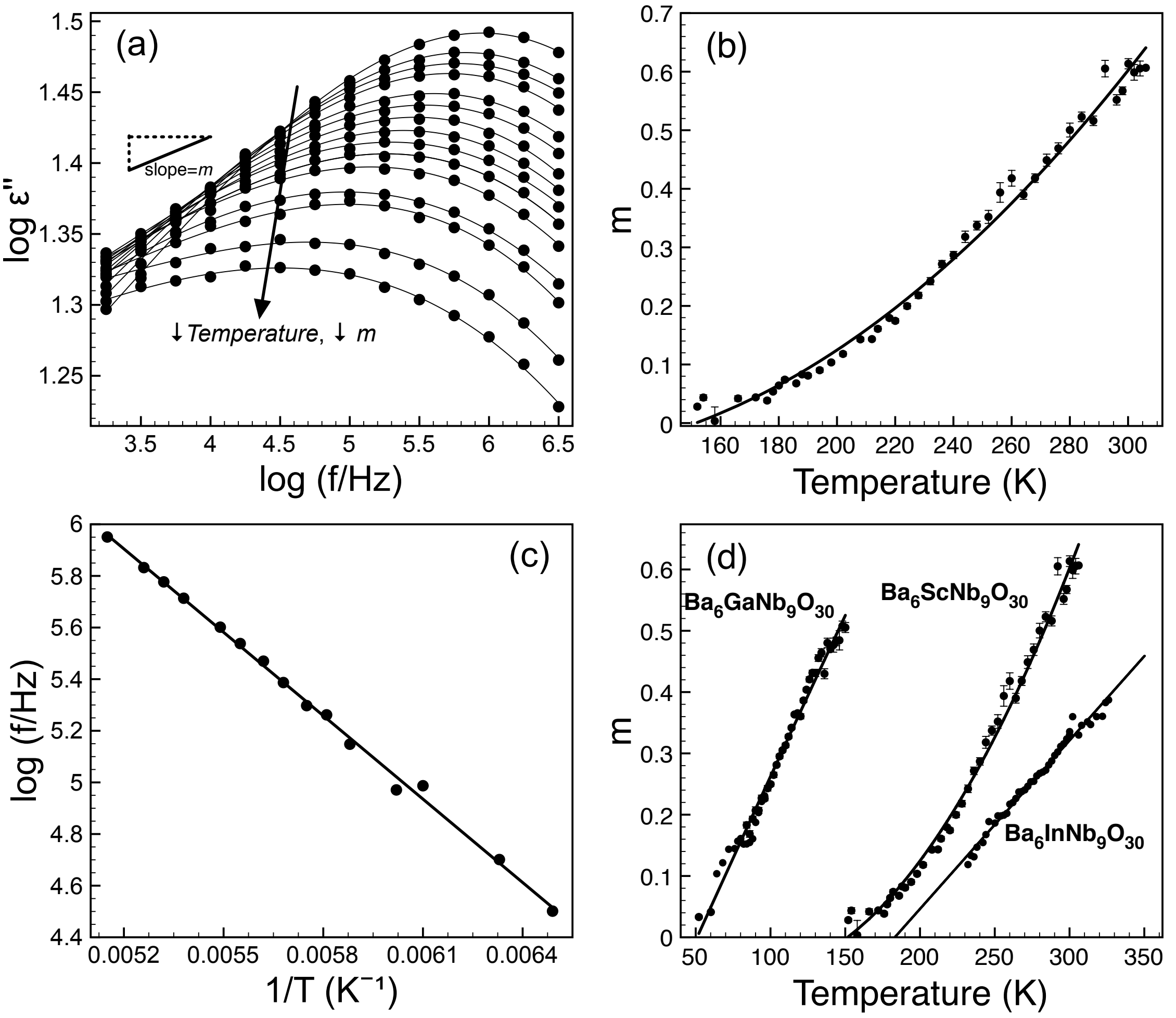

**Figure 4** Dielectric data for $Ba_6ScNB_9O_{30}$: loss peaks fitted to Jonscher's two exponent model (a); gradient, *m*, for $\varepsilon''(f)$ at $f < f_p$ (b); and Arrhenius plot for temperature dependence of $f_p$ (c). Comparison of *m* data for Ga, Sc and In analogues (d).

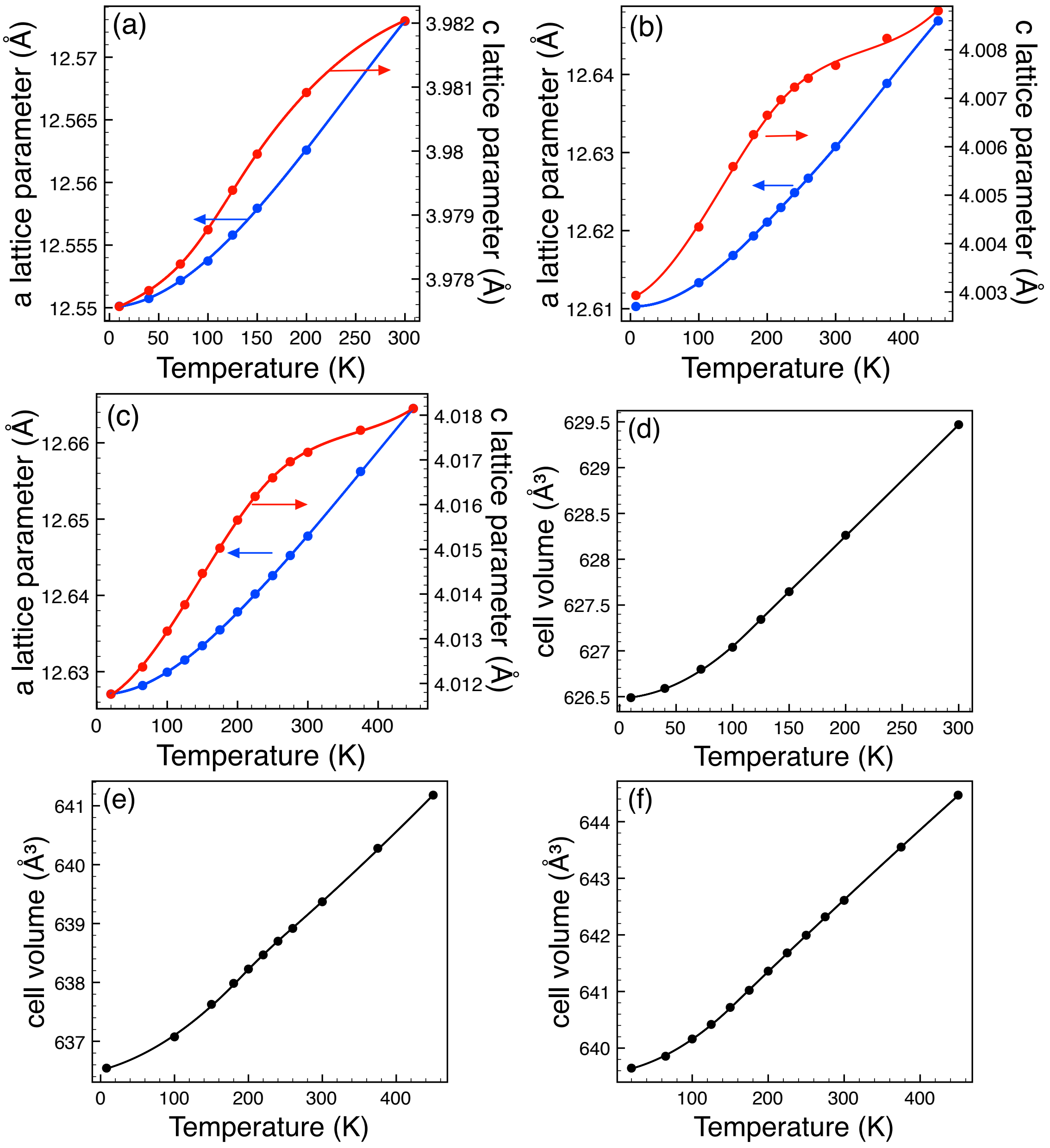

**Figure 5** (Colour online) Lattice parameters and unit cell volume as a function of temperature for $Ba_6GaNb_9O_{30}$ (a, d), $Ba_6ScNb_9O_{30}$ (b, e), and $Ba_6InNb_9O_{30}$ (c, f) from PND data.

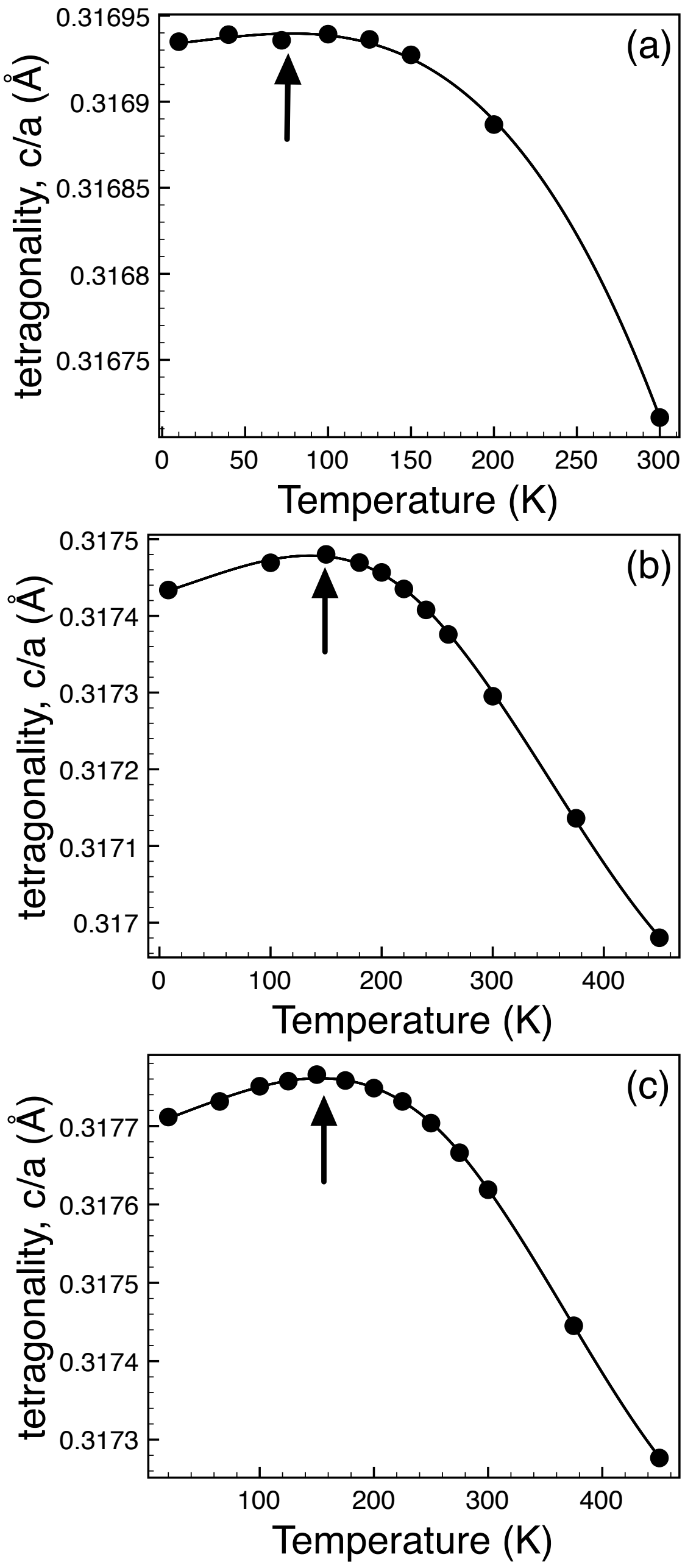

**Figure 6** Unit cell tetragonality, c/a, as a function of temperature for $Ba_6GaNb_9O_{30}$ (a), $Ba_6ScNb_9O_{30}$ (b), and $Ba_6InNb_9O_{30}$ (c).

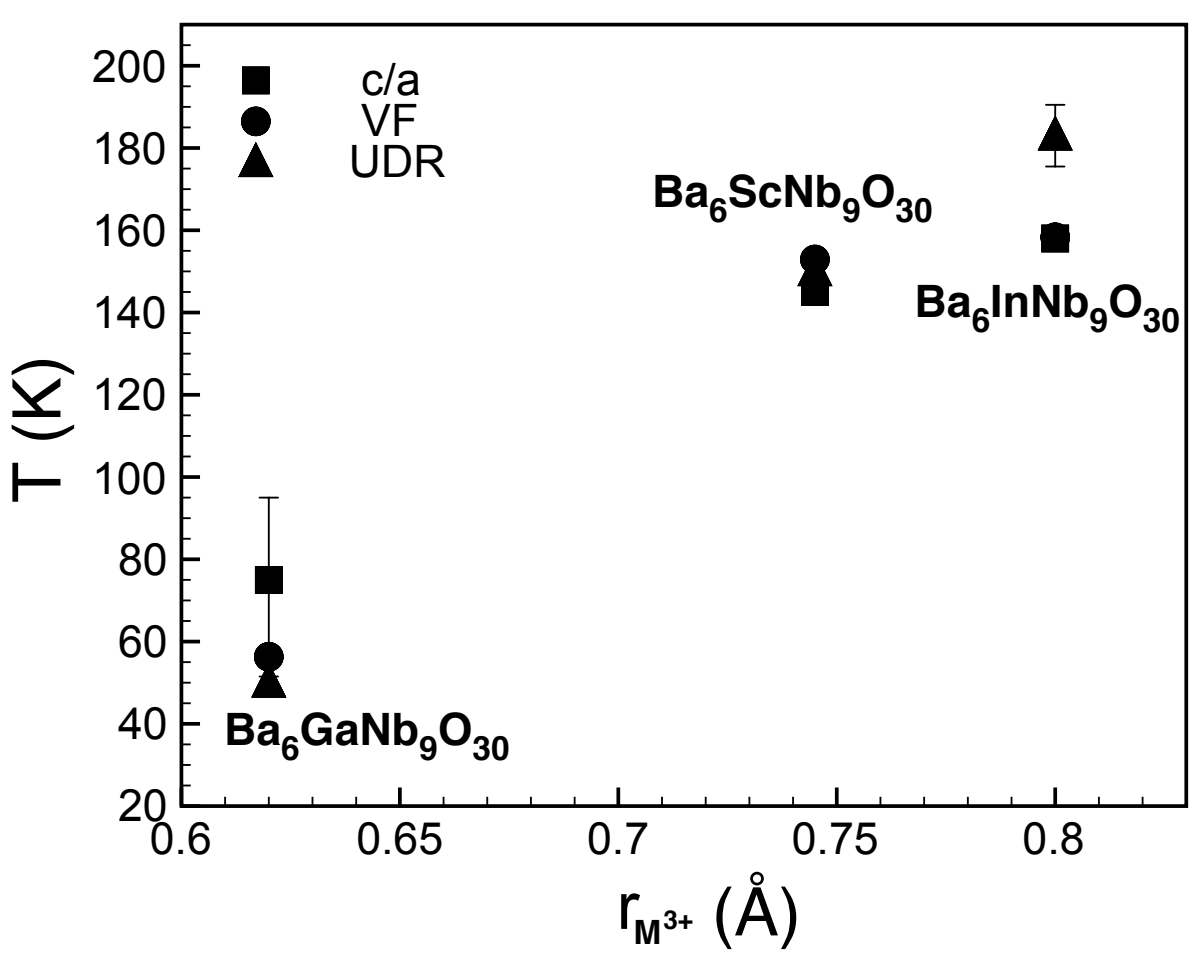

**Figure 7** Dependence of characteristic temperature parameters determined by the three different methods, as a function of $M^{3+}$ B-cation radius.

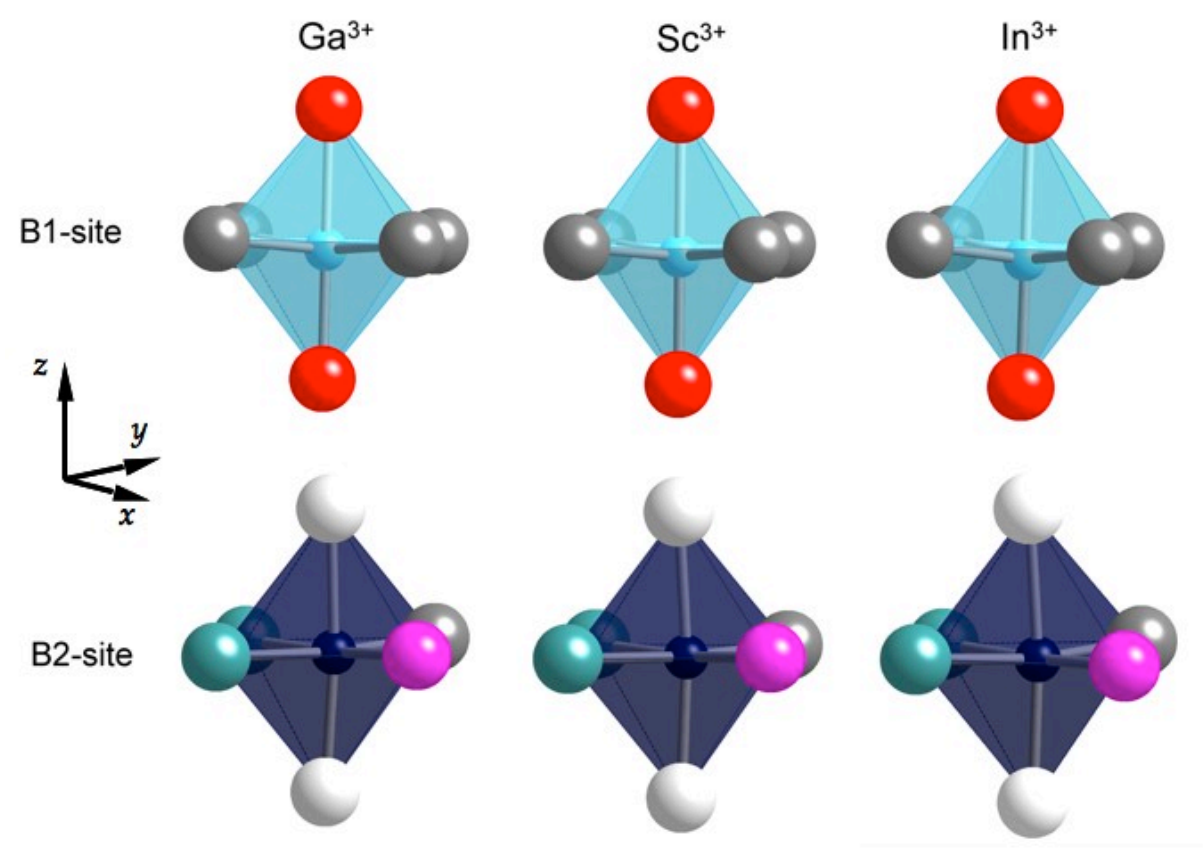

**Figure 8** (Colour online) B-cation displacements for both B1 and B2 octahedra in Ga, Sc and In analogues determined from refinement of data at 100 K in non-centrosymmetric space group *P4bm*.

**Origin and stability of the dipolar response in a family of tetragonal tungsten bronze relaxors**

Andrei Rotaru[1], Donna C. Arnold[1], Aziz Daoud-Aladine[2], Finlay D. Morrison[1*]

**Supplementary Information**

Tables of refinement parameters for $Ba_6GaNb_9O_{30}$, $Ba_6ScNb_9O_{30}$ and $Ba_6InNb_9O_{30}$ at various temperatures in the non-polar space group P4/mbm and the polar space group P4bm.

**Table S1:** Refinement details for $Ba_6GaNb_9O_{30}$ at all temperatures in the non-polar space group P4/mbm.

| Parameter | Temperature (K) | | | |
|---|---|---|---|---|
| | 10 | 40 | 72 | 100 |
| $\chi^2$ | 9.255 | 9.601 | 9.333 | 8.812 |
| wRp (%) | 5.11 | 5.20 | 5.12 | 4.97 |
| Rp (%) | 4.75 | 4.87 | 4.81 | 4.67 |
| a (Å) | 12.54973(3) | 12.55034(3) | 12.55186(3) | 12.55349(3) |
| c (Å) | 3.97735(2) | 3.97759(2) | 3.97804(2) | 3.97860(2) |
| Cell Volume (Å$^3$) | 626.415(3) | 626.514(3) | 626.737(3) | 626.987 (3) |
| Ba1 (0,0,0) | | | | |
| Ba1 U(iso) × 100 Å$^2$ | -0.09(4) | 0.02(4) | -0.07(4) | 0.04(4) |
| Ba2 (x,y,0) | 0.1712(1)<br>0.6712(1) | 0.1714(1)<br>0.6714(1) | 0.1717 (1)<br>0.6717(1) | 0.1715(1)<br>0.6715(1) |
| Ba2 U(iso) × 100 Å$^2$ | 0.73(4) | 0.79(4) | 0.76(4) | 0.90(4) |
| Nb1/M$^{3+}$1, (0,½, ½) | | | | |
| Nb1/ M$^{3+}$1 U(iso) × 100 Å$^2$ | 0.98(4) | 0.99(4) | 0.96(4) | 1.00(4) |
| Nb2/ M$^{3+}$2 (x,y, ½) | 0.07456(7)<br>0.21457(7) | 0.07466(7)<br>0.21454(7) | 0.07464(7)<br>0.21455(7) | 0.07457(7)<br>0.21451(7) |
| Nb2/ M$^{3+}$2 U(iso) × 100 Å$^2$ | 0.47(2) | 0.49(2) | 0.49(2) | 0.55(2) |
| O1 (0,½,0) | | | | |
| O1 U(iso) × 100 Å$^2$ | 1.34(5) | 1.36(5) | 1.28(5) | 1.24(5) |
| O2 (x,y, ½) | 0.28239(8)<br>0.78240(8) | 0.28235(8)<br>0.78235(8) | 0.28225(8)<br>0.78225(8) | 0.28229(8)<br>0.78229(8) |
| O2 U(iso) × 100 Å$^2$ | 0.40(3) | 0.42(4) | 0.34(3) | 0.43(3) |
| O3 (x,y,0) | 0.07672(9)<br>0.20766(8) | 0.07662(9)<br>0.20791(9) | 0.07684(9)<br>0.20768(8) | 0.07666(8)<br>0.20760(8) |
| O3 U(iso) × 100 Å$^2$ | 0.82(3) | 0.84(3) | 0.79(3) | 0.81(3) |
| O4 (x,y, ½) | 0.34378(8)<br>0.00713(7) | 0.34400(8)<br>0.00707(7) | 0.34380(8)<br>0.00706(7) | 0.34379(8)<br>0.00707(7) |
| O4 U(iso) × 100 Å$^2$ | 0.62(3) | 0.59(3) | 0.64(3) | 0.59(3) |
| O5 (x,y, ½) | 0.14102(8)<br>0.06941(8) | 0.14103(8)<br>0.06932(8) | 0.14120(8)<br>0.06936(8) | 0.14113(8)<br>0.06937(8) |
| O5 U(iso) × 100 Å$^2$ | 0.21(2) | 0.29(3) | 0.28(3) | 0.32(2) |

| | 125 | 150 | 200 | 300 |
|---|---|---|---|---|
| $\chi^2$ | 9.087 | 8.996 | 8.191 | 13.46 |
| wRp (%) | 5.07 | 5.03 | 4.80 | 6.19 |
| Rp (%) | 4.79 | 4.74 | 4.59 | 6.04 |
| a (Å) | 12.55539(2) | 12.55755(3) | 12.56220 (3) | 12.5723(3) |
| c (Å) | 3.97917(2) | 3.97973 (1) | 3.98071(2) | 3.98181(2) |
| Cell Volume (Å$^3$) | 627.267(3) | 627.573(3) | 628.191(3) | 629.384(4) |
| Ba1 (0,0,0) | | | | |
| Ba1 U(iso) × 100 Å$^2$ | 0.02(4) | 0.08(4) | 0.18(4) | 0.36(6) |
| Ba2 (x,y,0) | 0.1716(1)<br>0.6716(1) | 0.17158(10)<br>0.67158(10) | 0.1718(1)<br>0.6718(1) | 0.1722(1)<br>0.6722(1) |
| Ba2 U(iso) × 100 Å$^2$ | 1.00(4) | 1.02(4) | 1.18(4) | 1.56(5) |
| Nb1/M$^{3+}$1, (0,½, ½) | | | | |
| Nb1/ M$^{3+}$1 U(iso) × 100 Å$^2$ | 1.07(4) | 1.05(4) | 1.07(4) | 1.19(5) |
| Nb2/ M$^{3+}$2 (x,y, ½) | 0.07440(7)<br>0.21451(7) | 0.07453(7)<br>0.21452(7) | 0.07443(7)<br>0.21467(7) | 0.0743(1)<br>0.21481(9) |
| Nb2/ $M^{3+}$2 U(iso) × 100 Å$^2$ | 0.57 (2) | 0.61(2) | 0.67(2) | 0.76(3) |
| O1 (0,½,0) | | | | |
| O1 U(iso) × 100 Å$^2$ | 1.25(5) | 1.26(5) | 1.25(5) | 1.35(6) |
| O2 (x,y, ½) | 0.28227(8)<br>0.78228(8) | 0.28237(8)<br>0.78238(8) | 0.28226(8)<br>0.78226(8) | 0.2826(1)<br>0.7826(1) |
| O2 U(iso) × 100 Å$^2$ | 0.45(3) | 0.47(3) | 0.56(3) | 0.65(4) |
| O3 (x,y,0) | 0.07664(9)<br>0.20778(8) | 0.07676(9)<br>0.20773(8) | 0.07687(8)<br>0.20760(8) | 0.0768(1)<br>0.2078(1) |
| O3 U(iso) × 100 Å$^2$ | 0.87(3) | 0.81(3) | 0.80(3) | 0.90(3) |
| O4 (x,y, ½) | 0.34389(8)<br>0.00726(7) | 0.34375(8)<br>0.00719(7) | 0.34386(8)<br>0.00716(7) | 0.3436(1)<br>0.0070(1) |
| O4 U(iso) × 100 Å$^2$ | 0.68(3) | 0.69(3) | 0.71(2) | 0.94(4) |
| O5 (x,y, ½) | 0.14128(8)<br>0.06930(8) | 0.14117(8)<br>0.06935(8) | 0.14134(8)<br>0.06941(8) | 0.1416(1)<br>0.0696(1) |
| O5 U(iso) × 100 Å$^2$ | 0.34(3) | 0.37(3) | 0.41(2) | 0.62(3) |

**Table S2:** Refinement details for $Ba_6ScNb_9O_{30}$ at all temperatures in the non-polar space group P4/mbm.

| **Parameter** | **Temperature (K)** | | | |
|---|---|---|---|---|
| | **8** | **100** | **150** | **180** |
| $\chi^2$ | 12.24 | 11.47 | 10.92 | 10.17 |
| wRp (%) | 5.74 | 5.53 | 5.38 | 5.18 |
| Rp (%) | 5.12 | 4.98 | 4.90 | 4.79 |
| a (Å) | 12.60946(3) | 12.61246(3) | 12.61597(3) | 12.61852(3) |
| c (Å) | 4.00264(2) | 4.00404(2) | 4.00531(2) | 4.00598(1) |
| Cell Volume (Å$^3$) | 636.414(3) | 636.939(3) | 637.495(3) | 637.860(3) |
| Ba1 (0,0,0) | | | | |
| Ba1 U(iso) × 100 Å$^2$ | -0.36(4) | -0.32(4) | -0.22(4) | -0.12(4) |
| Ba2 (x,y,0) | 0.1710(1)<br>0.6710(1) | 0.1713(1)<br>0.6713(1) | 0.1712(1)<br>0.6712(1) | 0.1711(1)<br>0.6711(1) |
| Ba2 U(iso) × 100 Å$^2$ | 1.04(4) | 1.16(4) | 1.39(4) | 1.55(4) |
| Nb1/$M^{3+}$1, (0,½, ½) | | | | |
| Nb1/ $M^{3+}$1 U(iso) × 100 Å$^2$ | 1.34(4) | 1.37(4) | 1.53(4) | 1.56(4) |
| Nb2/ $M^{3+}$2 (x,y, ½) | 0.07531(6)<br>0.21430(6) | 0.07537(6)<br>0.21427(6) | 0.07534(6)<br>0.21442(6) | 0.07531(6)<br>0.21441(6) |
| Nb2/ $M^{3+}$2 U(iso) × 100 Å$^2$ | 0.44(2) | 0.47(2) | 0.56(23) | 0.61(2) |
| O1 (0,½,0) | | | | |
| O1 U(iso) × 100 Å$^2$ | 1.61(5) | 1.50(5) | 1.49(5) | 1.48(5) |
| O2 (x,y, ½) | 0.28151(8)<br>0.78151(8) | 0.28136(7)<br>0.78136(7) | 0.28134(8)<br>0.78134(8) | 0.28129(7)<br>0.78129(7) |
| O2 U(iso) × 100 Å$^2$ | 0.46(3) | 0.46(3) | 0.59(3) | 0.59(3) |
| O3 (x,y,0) | 0.07711(9)<br>0.20683(8) | 0.07726(8)<br>0.20672(8) | 0.07732(8)<br>0.20682(8) | 0.07724(8)<br>0.20691(8) |
| O3 U(iso) × 100 Å$^2$ | 1.06(3) | 1.01(3) | 1.03(3) | 1.07(3) |
| O4 (x,y, ½) | 0.34466(8)<br>0.00716(7) | 0.34474(8)<br>0.00706(7) | 0.34473(8)<br>0.00706(7) | 0.34460(8)<br>0.00699(7) |
| O4 U(iso) × 100 Å$^2$ | 0.86(3) | 0.83(3) | 0.86(3) | 0.92(3) |
| O5 (x,y, ½) | 0.14031(8)<br>0.06854(8) | 0.14036(8)<br>0.06850(8) | 0.14042(8)<br>0.06866(8) | 0.14058(8)<br>0.06875(8) |
| O5 U(iso) × 100 Å$^2$ | 0.56(3) | 0.55(3) | 0.66(3) | 0.69(3) |
| **Parameter** | **200** | **220** | **240** | **260** |
| $\chi^2$ | 10.46 | 10.08 | 9.792 | 9.404 |
| wRp (%) | 5.24 | 5.15 | 5.07 | 4.98 |
| Rp (%) | 4.78 | 4.75 | 4.62 | 4.56 |
| a (Å) | 12.62030(3) | 12.62217(3) | 12.62409(3) | 12.62603(3) |
| c (Å) | 4.00638(1) | 4.00671(1) | 4.00697(1) | 4.00717(2) |
| Cell Volume (Å$^3$) | 638.105(3) | 638.345(3) | 638.581(3) | 638.809(3) |
| Ba1 (0,0,0) | | | | |
| **Ba1 U(iso) × 100 Å$^2$** | -0.14(4) | -0.15(4) | -0.06(4) | 0.03(4) |
| **Ba2 (x,y,0)** | 0.1713(1)<br>0.6713(1) | 0.1713(1)<br>0.6713(1) | 0.1713(1)<br>0.6713(1) | 0.1714(1)<br>0.6714(1) |
| **Ba2 U(iso) × 100 Å$^2$** | 1.44(4) | 1.56(4) | 1.66(4) | 1.70(4) |
| **Nb1/M$^{3+}$1, (0,½, ½)** | | | | |
| **Nb1/ M$^{3+}$1 U(iso) × 100 Å$^2$** | 1.53(4) | 1.58(4) | 1.61(4) | 1.71(4) |
| **Nb2/ M$^{3+}$2 (x,y, ½)** | 0.07534(6)<br>0.21439(6) | 0.07525(6)<br>0.21448(6) | 0.07527(6)<br>0.21447(6) | 0.07522(6)<br>0.21457(6) |
| **Nb2/ M$^{3+}$2 U(iso) × 100 Å$^2$** | 0.54(2) | 0.58(2) | 0.59(2) | 0.68(2) |
| **O1 (0,½,0)** | | | | |
| **O1 U(iso) × 100 Å$^2$** | 1.51(5) | 1.45(5) | 1.50(5) | 1.45(5) |
| **O2 (x,y, ½)** | 0.28138(7)<br>0.78138(7) | 0.28121(7)<br>0.78121(7) | 0.28139(7)<br>0.78139(7) | 0.28129(7)<br>0.78129(7) |
| **O2 U(iso) × 100 Å$^2$** | 0.58(3) | 0.67(3) | 0.67(3) | 0.75(3) |
| **O3 (x,y,0)** | 0.07731(8)<br>0.20680(8) | 0.07730(8)<br>0.20686(8) | 0.07736(8)<br>0.20686(8) | 0.07725(8)<br>0.20685(8) |
| **O3 U(iso) × 100 Å$^2$** | 0.98(3) | 1.04(3) | 1.05(3) | 1.13(3) |
| **O4 (x,y, ½)** | 0.34474(8)<br>0.00693(7) | 0.34468(8)<br>0.00705(7) | 0.34468(8)<br>0.00696(7) | 0.34465(8)<br>0.00700(7) |
| **O4 U(iso) × 100 Å$^2$** | 0.81(3) | 0.93(3) | 0.94(3) | 1.05(3) |
| **O5 (x,y, ½)** | 0.14043(8)<br>0.06869(8) | 0.14060(8)<br>0.06876(8) | 0.14064(8)<br>0.06879(8) | 0.14078(7)<br>0.06887(8) |
| **O5 U(iso) × 100 Å$^2$** | 0.63(3) | 0.70(3) | 0.71(3) | 0.78(3) |

| | **300** | **375** | **450** | |
|---|---|---|---|---|
| **$\chi^2$** | 8.542 | 8.629 | 7.930 | |
| **wRp (%)** | 4.77 | 4.76 | 4.57 | |
| **Rp (%)** | 4.35 | 4.30 | 4.10 | |
| **a (Å)** | 12.63011(2) | 12.63817(2) | 12.64619(2) | |
| **c (Å)** | 4.00746(1) | 4.00801(1) | 4.00856(1) | |
| **Cell Volume (Å$^3$)** | 639.268(3) | 640.172(3) | 641.074(3) | |
| **Ba1 (0,0,0)** | | | | |
| **Ba1 U(iso) × 100 Å$^2$** | 0.04(4) | 0.19(4) | 0.26(4) | |
| **Ba2 (x,y,0)** | 0.1715(1)<br>0.6715(1) | 0.1716(1)<br>0.6716(1) | 0.1718(1)<br>0.6718(1) | |
| **Ba2 U(iso) × 100 Å$^2$** | 1.82(4) | 1.94(4) | 2.09(4) | |
| **Nb1/M$^{3+}$1, (0,½, ½)** | | | | |
| **Nb1/ M$^{3+}$1 U(iso) × 100 Å$^2$** | 1.72(4) | 1.81(4) | 1.80(4) | |
| **Nb2/ M$^{3+}$2 (x,y, ½)** | 0.07525(6)<br>0.21464(6) | 0.07522(6)<br>0.21482(6) | 0.07507(6)<br>0.21499(6) | |
| **Nb2/ M$^{3+}$2 U(iso) × 100 Å$^2$** | 0.68(2) | 0.72(2) | 0.74(2) | |
| **O1 (0,½,0)** | | | | |
| **O1 U(iso) × 100 Å$^2$** | 1.60(5) | 1.87(5) | 1.88(5) | |
| **O2 (x,y, ½)** | 0.28140(7)<br>0.78140(7) | 0.28154(8)<br>0.78154(8) | 0.28167(7)<br>0.78167(7) | |
| **O2 U(iso) × 100 Å$^2$** | 0.81(3) | 0.92(3) | 0.94(3) | |
| **O3 (x,y,0)** | 0.07741(8)<br>0.20693(8) | 0.07720(8)<br>0.20712(8) | 0.07703(8)<br>0.20713(8) | |
| **O3 U(iso) × 100 Å$^2$** | 1.16(2) | 1.27(3) | 1.37(3) | |
| **O4 (x,y, ½)** | 0.34476(8)<br>0.00708(7) | 0.34497(8)<br>0.00698(7) | 0.34507(8)<br>0.00703(7) | |
| **O4 U(iso) × 100 Å$^2$** | 1.09(3) | 1.18(3) | 1.29(3) | |
| **O5 (x,y, ½)** | 0.14075(7)<br>0.06891(8) | 0.14073(8)<br>0.06902(8) | 0.14098(8)<br>0.06920(8) | |
| **O5 U(iso) × 100 Å$^2$** | 0.83(2) | 0.92(3) | 1.03(3) | |

**Table S3:** Refinement details for $Ba_6InNb_9O_{30}$ at all temperatures in the non-polar space group P4/mbm.

| Parameter | Temperature (K) | | | |
|---|---|---|---|---|
| | 20 | 65 | 100 | 125 |
| $\chi^2$ | 10.09 | 9.806 | 9.744 | 9.018 |
| wRp (%) | 5.68 | 5.59 | 5.58 | 5.41 |
| Rp (%) | 4.89 | 4.83 | 4.86 | 4.73 |
| a (Å) | 12.62641(3) | 12.62754(3) | 12.62926(3) | 12.63093(3) |
| c (Å) | 4.01147(2) | 4.01209(2) | 4.01288(2) | 4.01348(2) |
| Cell Volume (Å$^3$) | 639.533(3) | 639.747(3) | 640.047(3) | 640.313(3) |
| Ba1 (0,0,0) | | | | |
| Ba1 U(iso) × 100 Å$^2$ | -0.31(4) | -0.28(4) | -0.24(4) | -0.22(4) |
| Ba2 (x,y,0) | 0.1710(1)<br>0.6710(1) | 0.1709(1)<br>0.6709(1) | 0.1709(1)<br>0.6709(1) | 0.1712(1)<br>0.6712(1) |
| Ba2 U(iso) × 100 Å$^2$ | 1.42(4) | 1.52(4) | 1.56(4) | 1.62(4) |
| Nb1/M$^{3+}$1, (0,½, ½) | | | | |
| Nb1/ M$^{3+}$1 U(iso) × 100 Å$^2$ | 0.89(4) | 0.92(4) | 0.91(4) | 0.94(4) |
| Nb2/ M$^{3+}$2 (x,y, ½) | 0.07579(7)<br>0.21412(7) | 0.07570(7)<br>0.21428(7) | 0.07572(7)<br>0.21422(7) | 0.07574(7)<br>0.21423(7) |
| Nb2/ M$^{3+}$2 U(iso) × 100 Å$^2$ | 0.56(3) | 0.62(3) | 0.61(3) | 0.65(3) |
| O1 (0,½,0) | | | | |
| O1 U(iso) × 100 Å$^2$ | 1.90(6) | 1.90(5) | 1.89(5) | 1.87(5) |
| O2 (x,y, ½) | 0.28105(8)<br>0.78105(8) | 0.28099(8)<br>0.78099(8) | 0.28094(8)<br>0.78095(8) | 0.28101(8)<br>0.78101(8) |
| O2 U(iso) × 100 Å$^2$ | 0.67(4) | 0.70(4) | 0.72(4) | 0.71(4) |
| O3 (x,y,0) | 0.07684(9)<br>0.20705(9) | 0.07711(9)<br>0.20698(9) | 0.07715(9)<br>0.20696(9) | 0.07715(9)<br>0.20695(9) |
| O3 U(iso) × 100 Å$^2$ | 1.34(3) | 1.36(3) | 1.30(3) | 1.33(3) |
| O4 (x,y, ½) | 0.34528(9)<br>0.00688(8) | 0.34517(9)<br>0.00673(8) | 0.34519(9)<br>0.00683(8) | 0.34516(8)<br>0.00677(8) |
| O4 U(iso) × 100 Å$^2$ | 0.98(3) | 1.04(3) | 1.01(3) | 1.04(3) |
| O5 (x,y, ½) | 0.14010(8)<br>0.06831(9) | 0.14013(8)<br>0.06840(9) | 0.14008(8)<br>0.06835(9) | 0.14022(8)<br>0.06832(9) |
| O5 U(iso) × 100 Å$^2$ | 0.76(3) | 0.78(3) | 0.75(3) | 0.80(3) |

| Parameter | 150 | 175 | 200 | 225 |
|---|---|---|---|---|
| $\chi^2$ | 8.545 | 8.200 | 8.043 | 7.754 |
| wRp (%) | 5.19 | 5.18 | 5.03 | 4.95 |
| Rp (%) | 4.57 | 4.63 | 4.41 | 4.38 |
| a (Å) | 12.63275(3) | 12.63489(3) | 12.63717(2) | 12.63951(2) |
| c (Å) | 4.01417(1) | 4.01475(1) | 4.01535(1) | 4.01588(1) |
| Cell Volume (Å$^3$) | 640.607(3) | 640.917(3) | 641.244(3) | 641.566(3) |
| Ba1 (0,0,0) | | | | |
| Ba1 U(iso) × 100 Å$^2$ | -0.17(4) | -0.12(4) | -0.15(4) | -0.07(4) |
| Ba2 (x,y,0) | 0.1713(1)<br>0.6713(1) | 0.1712(1)<br>0.6712(1) | 0.1712(1)<br>0.6712(1) | 0.1714(1)<br>0.6714(1) |
| Ba2 U(iso) × 100 Å$^2$ | 1.67(4) | 1.81(4) | 1.81(4) | 1.84(4) |
| Nb1/M$^{3+}$1, (0,½, ½) | | | | |
| Nb1/ M$^{3+}$1 U(iso) × 100 Å$^2$ | 1.02(4) | 1.06(4) | 1.04(4) | 1.09(4) |
| Nb2/ M$^{3+}$2 (x,y, ½) | 0.07577(7)<br>0.21424(7) | 0.07581(7)<br>0.21422(7) | 0.07580(7)<br>0.21423(7) | 0.07568(7)<br>0.21435(6) |
| Nb2/ M$^{3+}$2 U(iso) × 100 Å$^2$ | 0.71(3) | 0.71(3) | 0.68(3) | 0.71(3) |
| O1 (0,½,0) | | | | |
| O1 U(iso) × 100 Å$^2$ | 1.86(5) | 1.72(5) | 1.73(5) | 1.85(5) |
| O2 (x,y, ½) | 0.28098(8)<br>0.78098(8) | 0.28086(8)<br>0.78086(8) | 0.28093(7)<br>0.78093(7) | 0.07700(8)<br>0.20706(8) |
| O2 U(iso) × 100 Å$^2$ | 0.78(4) | 0.81(4) | 0.79(3) | 1.28(3) |
| O3 (x,y,0) | 0.07711(8)<br>0.20706(8) | 0.07706(8)<br>0.20706(8) | 0.07698(8)<br>0.20709(8) | 0.28089(7)<br>0.78089(7) |
| **O3 U(iso) × 100 Å²** | 1.35(3) | 1.34(3) | 1.31(3) | 0.86(3) |
| **O4 (x,y, ½)** | 0.34508(8)<br>0.00676(7) | 0.34519(8)<br>0.00675(7) | 0.34503(8)<br>0.00685(7) | 0.34513(8)<br>0.00671(7) |
| **O4 U(iso) × 100 Å²** | 1.08(3) | 1.09(3) | 1.06(3) | 1.12(3) |
| **O5 (x,y, ½)** | 0.14017(8)<br>0.06855(8) | 0.14030(8)<br>0.06845(8) | 0.14026(8)<br>0.06856(8) | 0.14035(8)<br>0.06852(8) |
| **O5 U(iso) × 100 Å²** | 0.84(3) | 0.84(3) | 0.84(3) | 0.91(3) |

| **Parameter** | **250** | **275** | **300** | **375** |
|---|---|---|---|---|
| $\chi^2$ | 7.691 | 7.401 | 7.248 | 6.540 |
| **wRp (%)** | 4.92 | 4.85 | 4.81 | 4.62 |
| **Rp (%)** | 4.38 | 4.41 | 4.31 | 4.12 |
| **a (Å)** | 12.64195(2) | 12.64450(2) | 12.64713(2) | 12.65544(2) |
| **c (Å)** | 4.01631(1) | 4.01665(1) | 4.01688(1) | 4.01736(1) |
| **Cell Volume (Å³)** | 641.883(3) | 642.196(3) | 642.500(3) | 643.420(3) |
| **Ba1 (0,0,0)** | | | | |
| **Ba1 U(iso) × 100 Å²** | -0.06(4) | -0.02(4) | 0.04(4) | 0.24(4) |
| **Ba2 (x,y,0)** | 0.1714(1)<br>0.6714(1) | 0.1715(1)<br>0.6715(1) | 0.1715(1)<br>0.6715(1) | 0.1718(1)<br>0.6718(1) |
| **Ba2 U(iso) × 100 Å²** | 1.93(4) | 2.02(4) | 2.06(4) | 2.32(4) |
| **Nb1/$M^{3+}$1, (0,½, ½)** | | | | |
| **Nb1/ $M^{3+}$1 U(iso) × 100 Å²** | 1.15(4) | 1.19(4) | 1.20(4) | 1.42(4) |
| **Nb2/ $M^{3+}$2 (x,y, ½)** | 0.07570(7)<br>0.21440(7) | 0.07564(7)<br>0.21447(6) | 0.07571(7)<br>0.21442(6) | 0.07547(7)<br>0.21467(6) |
| **Nb2/ $M^{3+}$2 U(iso) × 100 Å²** | 0.76(3) | 0.79(3) | 0.77(3) | 0.87(2) |
| **O1 (0,½,0)** | | | | |
| **O1 U(iso) × 100 Å²** | 1.81(5) | 1.81(5) | 1.78(5) | 2.04(5) |
| **O2 (x,y, ½)** | 0.28098(7)<br>0.78098(7) | 0.28088(7)<br>0.78088(7) | 0.28106(8)<br>0.78106(8) | 0.28115(8)<br>0.78115(8) |
| **O2 U(iso) × 100 Å²** | 0.88(3) | 0.95(3) | 0.94(3) | 1.19(4) |
| **O3 (x,y,0)** | 0.07708(8)<br>0.20712(8) | 0.07705(8)<br>0.20698(8) | 0.07694(8)<br>0.20707(8) | 0.07698(8)<br>0.20704(8) |
| **O3 U(iso) × 100 Å²** | 1.35(3) | 1.38(3) | 1.38(3) | 1.55(3) |
| **O4 (x,y, ½)** | 0.34514(8)<br>0.00690(7) | 0.34510(8)<br>0.00685(7) | 0.34505(8)<br>0.00683(7) | 0.34530(8)<br>0.00674(7) |
| **O4 U(iso) × 100 Å²** | 1.15(3) | 1.21(3) | 1.25(3) | 1.39(3) |
| **O5 (x,y, ½)** | 0.14047(8)<br>0.06862(8) | 0.14055(8)<br>0.06863(8) | 0.14062(8)<br>0.06864(8) | 0.14076(8)<br>0.06889(8) |
| **O5 U(iso) × 100 Å²** | 0.91(3) | 0.98(3) | 1.02(3) | 1.15(3) |
| **Parameter** | **450** | | | |
| $\chi^2$ | 6.396 | | | |
| **wRp (%)** | 4.52 | | | |
| **Rp (%)** | 4.04 | | | |
| **a (Å)** | 12.66382(2) | | | |
| **c (Å)** | 4.01788(1) | | | |
| **Cell Volume (Å³)** | 644.358(3) | | | |
| **Ba1 (0,0,0)** | | | | |
| **Ba1 U(iso) × 100 Å²** | 0.23(4) | | | |
| **Ba2 (x,y,0)** | 0.1719(1)<br>0.6719(1) | | | |
| **Ba2 U(iso) × 100 Å²** | 2.39(4) | | | |
| **Nb1/$M^{3+}$1, (0,½, ½)** | | | | |
| **Nb1/ $M^{3+}$1 U(iso) × 100 Å²** | 1.35(4) | | | |
| **Nb2/ $M^{3+}$2 (x,y, ½)** | 0.07532(7)<br>0.21483(6) | | | |
| **Nb2/ $M^{3+}$2 U(iso) × 100 Å²** | 0.86(3) | | | |
| **O1 (0,½,0)** | | | | |
| **O1 U(iso) × 100 Å$^2$** | 2.18(5) | | | |
| **O2 (x,y, ½)** | 0.28149(8)<br>0.78149(8) | | | |
| **O2 U(iso) × 100 Å$^2$** | 1.21(4) | | | |
| **O3 (x,y,0)** | 0.07674(8)<br>0.20714(8) | | | |
| **O3 U(iso) × 100 Å$^2$** | 1.63(3) | | | |
| **O4 (x,y, ½)** | 0.34520(8)<br>0.00682(7) | | | |
| **O4 U(iso) × 100 Å$^2$** | 1.49(3) | | | |
| **O5 (x,y, ½)** | 0.14084(8)<br>0.06895(8) | | | |
| **O5 U(iso) × 100 Å$^2$** | 1.25(3) | | | |

**Table S4:** Refinement details for $Ba_6GaNb_9O_{30}$ at all temperatures in the non-polar space group P4/mbm with the B-site ion thermal parameters set as Uaniso.

| **Parameter** | **Temperature (K)** | | | |
|---|---|---|---|---|
| | **10** | **40** | **72** | **100** |
| $\chi^2$ | 8.929 | 9.196 | 8.973 | 8.526 |
| **wRp (%)** | 5.02 | 5.09 | 5.02 | 4.89 |
| **Rp (%)** | 4.71 | 4.81 | 4.78 | 4.63 |
| **a (Å)** | 12.54975(3) | 12.55036(3) | 12.55188 (3) | 12.55351(3) |
| **c (Å)** | 3.97736(2) | 3.97761(2) | 3.97805(2) | 3.97861(2) |
| **Cell Volume (Å$^3$)** | 626.419(3) | 626.519(4) | 626.741(3) | 626.992(3) |
| **Ba1 (0,0,0)** | | | | |
| **Ba1 U(iso) × 100 Å$^2$** | -0.02(4) | 0.08(4) | -0.01(4) | 0.10(4) |
| **Ba2 (x,y,0)** | 0.1712(1)<br>0.6712(1) | 0.1714(1)<br>0.6714(1) | 0.1717(1)<br>0.6717(1) | 0.1714(1)<br>0.6714(1) |
| **Ba2 U(iso) × 100 Å$^2$** | 0.82(4) | 0.88(4) | 0.85(4) | 1.01(4) |
| **Nb1/M$^{3+}$1, (0,½, ½)** | | | | |
| **Nb1/ M$^{3+}$1 U(aniso) × 100 Å$^2$**<br>**$U_{11}$**<br>**$U_{22}$**<br>**$U_{33}$**<br>**$U_{12}$**<br>**$U_{13}$**<br>**$U_{23}$** | <br>0.46(5)<br>0.46(5)<br>1.9(1)<br>-0.24(7)<br>0.0<br>0.0 | <br>0.46(5)<br>0.46(5)<br>1.9(1)<br>-0.24(7)<br>0.0<br>0.0 | <br>0.39(5)<br>0.39(5)<br>2.0(1)<br>-0.11(7)<br>0.0<br>0.0 | <br>0.43(5)<br>0.43(5)<br>2.1(1)<br>-0.07(7)<br>0.0<br>0.0 |
| **Nb2/ M$^{3+}$2 (x,y, ½)** | 0.07470(7)<br>0.21453(7) | 0.07481(7)<br>0.21450(7) | 0.07480(7)<br>0.21451(7) | 0.07472(7)<br>0.21450(6) |
| **Nb2/ M$^{3+}$2 U(aniso) × 100 Å$^2$**<br>**$U_{11}$**<br>**$U_{22}$**<br>**$U_{33}$**<br>**$U_{12}$**<br>**$U_{13}$**<br>**$U_{23}$** | <br>0.54(4)<br>0.06(4)<br>0.90(4)<br>0.07(4)<br>0.0<br>0.0 | <br>0.53(4)<br>0.06(4)<br>0.99(4)<br>0.12(4)<br>0.0<br>0.0 | <br>0.51(4)<br>0.11(4)<br>0.93(4)<br>0.14(4)<br>0.0<br>0.0 | <br>0.64(4)<br>0.13(4)<br>1.00(4)<br>0.11(4)<br>0.0<br>0.0 |
| **O1 (0,½,0)** | | | | |
| **O1 U(iso) × 100 Å$^2$** | 1.28(5) | 1.30(5) | 1.22(5) | 1.18(5) |
| **O2 (x,y, ½)** | 0.28226(8)<br>0.78226(8) | 0.28221(8)<br>0.78221(8) | 0.28213(8)<br>0.78214(8) | 0.28216(8)<br>0.78217(8) |
| **O2 U(iso) × 100 Å$^2$** | 0.38(3) | 0.41(4) | 0.33(3) | 0.42(3) |
| **O3 (x,y,0)** | 0.07673(8)<br>0.20749(8) | 0.07664(9)<br>0.20773(8) | 0.07685(8)<br>0.20752(8) | 0.07668(8)<br>0.20741(8) |
| **O3 U(iso) × 100 Å$^2$** | 0.80(3) | 0.82(3) | 0.78(3) | 0.80(3) |
| **O4 (x,y, ½)** | 0.34389(8)<br>0.00719(7) | 0.34412(9)<br>0.00711(7) | 0.34387(9)<br>0.00710(7) | 0.34390(8)<br>0.00715(7) |
| **O4 U(iso) × 100 Å$^2$** | 0.58(3) | 0.56(3) | 0.61(3) | 0.57(3) |
| **O5 (x,y, ½)** | 0.14100(8) | 0.14099(8) | 0.14119(8) | 0.14115(8) |
| | 0.06926(8) | 0.06914(8) | 0.06918(8) | 0.06920(8) |
| **O5 U(iso) × 100 Å$^2$** | 0.30(3) | 0.38(3) | 0.36(3) | 0.41(2) |
| **Parameter** | **125** | **150** | **200** | **300** |
| **$\chi^2$** | 8.687 | 8.684 | 7.845 | 13.20 |
| **wRp (%)** | 4.96 | 4.95 | 4.70 | 6.13 |
| **Rp (%)** | 4.74 | 4.70 | 4.55 | 6.01 |
| **a (Å)** | 12.55541(3) | 12.55757(3) | 12.56222(3) | 12.57240(3) |
| **c (Å)** | 3.97918(2) | 3.97975(2) | 3.98072(2) | 3.98182(2) |
| **Cell Volume (Å$^3$)** | 627.272(3) | 627.577(3) | 628.195(3) | 629.387(4) |
| **Ba1 (0,0,0)** | | | | |
| **Ba1 U(iso) × 100 Å$^2$** | 0.09(4) | 0.15(4) | 0.23(4) | 0.40(6) |
| **Ba2 (x,y,0)** | 0.1715(1)<br>0.6715(1) | 0.1715(1)<br>0.6715(1) | 0.1717(1)<br>0.6717(1) | 0.1722(1)<br>0.6722(1) |
| **Ba2 U(iso) × 100 Å$^2$** | 1.11(4) | 1.14(4) | 1.29(4) | 1.65(5) |
| **Nb1/M$^{3+}$1, (0,½, ½)** | | | | |
| **Nb1/ M$^{3+}$1 U(aniso) × 100 Å$^2$**<br>**$U_{11}$**<br>**$U_{22}$**<br>**$U_{33}$**<br>**$U_{12}$**<br>**$U_{13}$**<br>**$U_{23}$** | <br>0.46(5)<br>0.46(5)<br>2.2(1)<br>-0.17(7)<br>0.0<br>0.0 | <br>0.48(5)<br>0.48(5)<br>2.1(1)<br>0.01(7)<br>0.0<br>0.0 | <br>0.50(5)<br>0.50(5)<br>2.14(9)<br>-0.02(7)<br>0.0<br>0.0 | <br>0.78(7)<br>0.78(7)<br>2.0(1)<br>-0.12(9)<br>0.0<br>0.0 |
| **Nb2/ M$^{3+}$2 (x,y, ½)** | 0.07454(7)<br>0.21447(7) | 0.07465(8)<br>0.21452(7) | 0.07458(7)<br>0.21466(6) | 0.0745(1)<br>0.21479(9) |
| **Nb2/ M$^{3+}$2 U(aniso) × 100 Å$^2$**<br>**$U_{11}$**<br>**$U_{22}$**<br>**$U_{33}$**<br>**$U_{12}$**<br>**$U_{13}$**<br>**$U_{23}$** | <br>0.66(4)<br>0.15(4)<br>1.03(4)<br>0.13(4)<br>0.0<br>0.0 | <br>0.73(4)<br>0.20(4)<br>1.02(4)<br>0.08(4)<br>0.0<br>0.0 | <br>0.69(4)<br>0.29(4)<br>1.13(4)<br>0.14(4)<br>0.0<br>0.0 | <br>0.85(6)<br>0.34(5)<br>1.21(5)<br>0.13(5)<br>0.0<br>0.0 |
| **O1 (0,½,0)** | | | | |
| **O1 U(iso) × 100 Å$^2$** | 1.18(5) | 1.20(5) | 1.18(4) | 1.28(6) |
| **O2 (x,y, ½)** | 0.28215(8)<br>0.78215(8) | 0.28225(8)<br>0.78225(8) | 0.28215(8)<br>0.78215(8) | 0.2825(1)<br>0.7825(1) |
| **O2 U(iso) × 100 Å$^2$** | 0.45(3) | 0.46(3) | 0.55(3) | 0.66(5) |
| **O3 (x,y,0)** | 0.07665(8)<br>0.20758(8) | 0.07678(8)<br>0.20754(8) | 0.07690(8)<br>0.20743(8) | 0.0768(1)<br>0.2076(1) |
| **O3 U(iso) × 100 Å$^2$** | 0.85(3) | 0.80(3) | 0.79(3) | 0.88(3) |
| **O4 (x,y, ½)** | 0.34401(8)<br>0.00731(7) | 0.34387(9)<br>0.00727(7) | 0.34393(8)<br>0.00722(7) | 0.3438(1)<br>0.0071(1) |
| **O4 U(iso) × 100 Å$^2$** | 0.65(3) | 0.66(3) | 0.68(3) | 0.92(4) |
| **O5 (x,y, ½)** | 0.14128(8)<br>0.06913(8) | 0.14122(8)<br>0.06919(8) | 0.14135(8)<br>0.06924(8) | 0.1416(1)<br>0.0695(1) |
| **O5 U(iso) × 100 Å$^2$** | 0.43(3) | 0.45(3) | 0.51(2) | 0.71(3) |

**Table S5:** Refinement details for $Ba_6ScNb_9O_{30}$ at all temperatures in the non-polar space group P4/mbm with the B-site ion thermal parameters set as Uaniso.

| **Parameter** | **Temperature (K)** | | | |
|---|---|---|---|---|
| | **8** | **100** | **150** | **180** |
| **$\chi^2$** | 11.16 | 10.49 | 9.947 | 9.337 |
| **wRp (%)** | 5.48 | 5.27 | 5.14 | 4.96 |
| **Rp (%)** | 4.97 | 4.85 | 4.74 | 4.64 |
| **a (Å)** | 12.60950(3) | 12.61249(3) | 12.61600(3) | 12.61855(3) |
| **c (Å)** | 4.0026(1) | 4.00406(1) | 4.00532(1) | 4.00600(1) |
| **Cell Volume (Å$^3$)** | 636.421(3) | 636.945(3) | 637.501(3) | 637.866(3) |
| **Ba1 (0,0,0)** | | | | |
| **Ba1 U(iso) × 100 Å$^2$** | -0.27(4) | -0.23(4) | -0.13(4) | -0.04(4) |
| **Ba2 (x,y,0)** | 0.1709(1)<br>0.6709(1) | 0.1711(1)<br>0.6711(1) | 0.1711(1)<br>0.6711(1) | 0.1710(1)<br>0.6710(1) |
| **Ba2 U(iso) × 100 Å$^2$** | 1.18(4) | 1.31(4) | 1.53(4) | 1.67(4) |
| **Nb1/M$^{3+}$1, (0,½, ½)** | | | | |
| **Nb1/ M$^{3+}$1 U(aniso) × 100 Å$^2$**<br>**$U_{11}$**<br>**$U_{22}$**<br>**$U_{33}$**<br>**$U_{12}$**<br>**$U_{13}$**<br>**$U_{23}$** | <br>0.33(5)<br>0.33(5)<br>3.20(9)<br>-0.15(6)<br>0.0<br>0.0 | <br>0.40(5)<br>0.40(5)<br>3.18(9)<br>-0.03(6)<br>0.0<br>0.0 | <br>0.61(5)<br>0.61(5)<br>3.3(1)<br>-0.05(2)<br>0.0<br>0.0 | <br>0.68(5)<br>0.68(5)<br>3.28(9)<br>-0.15(6)<br>0.0<br>0.0 |
| **Nb2/ M$^{3+}$2 (x,y, ½)** | 0.07554(6)<br>0.21421(6) | 0.07556(6)<br>0.21421(5) | 0.07554(6)<br>0.21436(6) | 0.07552(6)<br>0.21432(6) |
| **Nb2/ M$^{3+}$2 U(aniso) × 100 Å$^2$**<br>**$U_{11}$**<br>**$U_{22}$**<br>**$U_{33}$**<br>**$U_{12}$**<br>**$U_{13}$**<br>**$U_{23}$** | <br>0.44(4)<br>0.01(4)<br>0.89(4)<br>0.13(3)<br>0.0<br>0.0 | <br>0.43(4)<br>0.08(4)<br>0.92(3)<br>0.12(3)<br>0.0<br>0.0 | <br>0.57(4)<br>0.22(4)<br>0.94(4)<br>0.16(3)<br>0.0<br>0.0 | <br>0.61(4)<br>0.32(4)<br>0.96(3)<br>0.16(3)<br>0.0<br>0.0 |
| **O1 (0,½,0)** | | | | |
| **O1 U(iso) × 100 Å$^2$** | 1.54(5) | 1.43(5) | 1.43(5) | 1.43(4) |
| **O2 (x,y, ½)** | 0.28131(7)<br>0.78131(7) | 0.28118(7)<br>0.78118(7) | 0.28119(7)<br>0.78119(7) | 0.28117(7)<br>0.78117(7) |
| **O2 U(iso) × 100 Å$^2$** | 0.45(3) | 0.45(3) | 0.60(3) | 0.59(3) |
| **O3 (x,y,0)** | 0.07701(8)<br>0.20671(8) | 0.07717(8)<br>0.20662(8) | 0.07723(8)<br>0.20671(8) | 0.07716(8)<br>0.20681(8) |
| **O3 U(iso) × 100 Å$^2$** | 1.05(3) | 1.01(3) | 1.05(3) | 1.08(3) |
| **O4 (x,y, ½)** | 0.34476(8)<br>0.00725(7) | 0.34481(8)<br>0.00714(7) | 0.34481(8)<br>0.00713(7) | 0.34466(8)<br>0.00705(7) |
| **O4 U(iso) × 100 Å$^2$** | 0.81(3) | 0.79(3) | 0.83(3) | 0.90(3) |
| **O5 (x,y, ½)** | 0.14028(8)<br>0.06829(8) | 0.14036(8)<br>0.06827(8) | 0.14043(8)<br>0.06847(8) | 0.14056(8)<br>0.06856(8) |
| **O5 U(iso) × 100 Å$^2$** | 0.65(3) | 0.64(2) | 0.73(3) | 0.76(2) |
| **Parameter** | **200** | **220** | **240** | **260** |
| **$\chi^2$** | 9.593 | 9.216 | 8.841 | 8.513 |
| **wRp (%)** | 5.02 | 4.92 | 4.82 | 4.74 |
| **Rp (%)** | 4.63 | 4.60 | 4.46 | 4.41 |
| **a (Å)** | 12.62034(3) | 12.62220(3) | 12.62412(3) | 12.62606(2) |
| **c (Å)** | 4.00640(1) | 4.00673(1) | 4.00699(1) | 4.00719(1) |
| **Cell Volume (Å$^3$)** | 638.111(3) | 638.351(3) | 638.588(3) | 638.816(3) |
| **Ba1 (0,0,0)** | | | | |
| **Ba1 U(iso) × 100 Å$^2$** | -0.04(4) | -0.05(4) | 0.04(4) | 0.13(4) |
| **Ba2 (x,y,0)** | 0.1711(1)<br>0.6711(1) | 0.1711(1)<br>0.6711(1) | 0.1711(1)<br>0.6711(1) | 0.1712(1)<br>0.6712(1) |
| **Ba2 U(iso) × 100 Å$^2$** | 1.58(4) | 1.70(4) | 1.81(4) | 1.84(4) |
| **Nb1/M$^{3+}$1, (0,½, ½)** | | | | |
| **Nb1/ M$^{3+}$1 U(aniso) × 100 Å$^2$**<br>**$U_{11}$**<br>**$U_{22}$**<br>**$U_{33}$**<br>**$U_{12}$**<br>**$U_{13}$**<br>**$U_{23}$** | <br>0.58(5)<br>0.58(5)<br>3.5(1)<br>0.08(6)<br>0.0<br>0.0 | <br>0.64(5)<br>0.64(5)<br>3.5(1)<br>0.05(6)<br>0.0<br>0.0 | <br>0.65(5)<br>0.65(5)<br>3.50(9)<br>-0.03(6)<br>0.0<br>0.0 | <br>0.73(5)<br>0.73(5)<br>3.63(9<br>-0.09(6)<br>0.0<br>0.0 |
| **Nb2/ M$^{3+}$2 (x,y, ½)** | 0.07554(6)<br>0.21435(6) | 0.07545(6)<br>0.21444(6) | 0.07547(6)<br>0.21443(5) | 0.07544(6)<br>0.21450(5) |
| **Nb2/ $M^{3+}$2 U(aniso) × 100 Å$^2$** <br> **$U_{11}$** <br> **$U_{22}$** <br> **$U_{33}$** <br> **$U_{12}$** <br> **$U_{13}$** <br> **$U_{23}$** | <br> 0.64(4) <br> 0.19(4) <br> 0.86(3) <br> 0.18(3) <br> 0.0 <br> 0.0 | <br> 0.58(4) <br> 0.31(4) <br> 0.91(3) <br> 0.13(3) <br> 0.0 <br> 0.0 | <br> 0.66(4) <br> 0.23(4) <br> 0.96(3) <br> 0.15(3) <br> 0.0 <br> 0.0 | <br> 0.70(4) <br> 0.36(4) <br> 1.05(3) <br> 0.19(3) <br> 0.0 <br> 0.0 |
| **O1 (0,½,0)** | | | | |
| **O1 U(iso) × 100 Å$^2$** | 1.44(5) | 1.40(5) | 1.44(4) | 1.40(4) |
| **O2 (x,y, ½)** | 0.28126(7) <br> 0.78126(7) | 0.28108(7) <br> 0.78108(7) | 0.28124(7) <br> 0.78124(7) | 0.28117(7) <br> 0.78117(7) |
| **O2 U(iso) × 100 Å$^2$** | 0.59(3) | 0.68(3) | 0.68(3) | 0.75(3) |
| **O3 (x,y,0)** | 0.07725(8) <br> 0.20665(8) | 0.07723(8) <br> 0.20675(8) | 0.07729(8) <br> 0.20671(7) | 0.07717(7) <br> 0.20672(7) |
| **O3 U(iso) × 100 Å$^2$** | 0.99(3) | 1.05(3) | 1.06(2) | 1.14(2) |
| **O4 (x,y, ½)** | 0.34484(8) <br> 0.00702(7) | 0.34473(8) <br> 0.00713(7) | 0.34478(8) <br> 0.00706(7) | 0.34470(8) <br> 0.00708(7) |
| **O4 U(iso) × 100 Å$^2$** | 0.79(3) | 0.91(3) | 0.91(3) | 1.03(3) |
| **O5 (x,y, ½)** | 0.14049(8) <br> 0.06853(8) | 0.14065(8) <br> 0.06860(8) | 0.14069(8) <br> 0.06861(8) | 0.14080(7) <br> 0.06867(8) |
| **O5 U(iso) × 100 Å$^2$** | 0.70(3) | 0.77(2) | 0.80(2) | 0.86(2) |
| **Parameter** | **300** | **375** | **450** | |
| **$\chi^2$** | 7.794 | 7.906 | 7.332 | |
| **wRp (%)** | 4.55 | 4.56 | 4.39 | |
| **Rp (%)** | 4.24 | 4.20 | 3.99 | |
| **a (Å)** | 12.63014(2) | 12.63820(2) | 12.64623(2) | |
| **c (Å)** | 4.00748(1) | 4.00803(1) | 4.00858(1) | |
| **Cell Volume (Å$^3$)** | 639.275(3) | 640.179(3) | 641.081(3) | |
| **Ba1 (0,0,0)** | | | | |
| **Ba1 U(iso) × 100 Å$^2$** | 0.13(4) | 0.29(4) | 0.37(4) | |
| **Ba2 (x,y,0)** | 0.1713(1) <br> 0.6713(1) | 0.1715(1) <br> 0.6715(1) | 0.1717(1) <br> 0.6717(1) | |
| **Ba2 U(iso) × 100 Å$^2$** | 1.95(4) | 2.08(4) | 2.24(4) | |
| **Nb1/$M^{3+}$1, (0,½, ½)** | | | | |
| **Nb1/ $M^{3+}$1 U(aniso) × 100 Å$^2$** <br> **$U_{11}$** <br> **$U_{22}$** <br> **$U_{33}$** <br> **$U_{12}$** <br> **$U_{13}$** <br> **$U_{23}$** | <br> 0.77(5) <br> 0.77(5) <br> 3.57(9) <br> 0.00(6) <br> 0.0 <br> 0.0 | <br> 0.92(5) <br> 0.92(5) <br> 3.6(1) <br> -0.03(6) <br> 0.0 <br> 0.0 | <br> 0.93(5) <br> 0.93(5) <br> 3.49(9) <br> -0.07(6) <br> 0.0 <br> 0.0 | |
| **Nb2/ $M^{3+}$2 (x,y, ½)** | 0.07545(6) <br> 0.21459(5) | 0.07539(6) <br> 0.21476(5) | 0.07524(6) <br> 0.21494(5) | |
| **Nb2/ $M^{3+}$2 U(aniso) × 100 Å$^2$** <br> **$U_{11}$** <br> **$U_{22}$** <br> **$U_{33}$** <br> **$U_{12}$** <br> **$U_{13}$** <br> **$U_{23}$** | <br> 0.71(4) <br> 0.37(4) <br> 1.04(3) <br> 0.17(3) <br> 0.0 <br> 0.0 | <br> 0.80(4) <br> 0.35(4) <br> 1.12(3) <br> 0.16(3) <br> 0.0 <br> 0.0 | <br> 0.86(4) <br> 0.33(4) <br> 1.17(3) <br> 0.16(3) <br> 0.0 <br> 0.0 | |
| **O1 (0,½,0)** | | | | |
| **O1 U(iso) × 100 Å$^2$** | 1.55(4) | 1.80(5) | 1.80(5) | |
| **O2 (x,y, ½)** | 0.28127(7) <br> 0.78127(7) | 0.28142(7) <br> 0.78142(7) | 0.28156(7) <br> 0.78156(7) | |
| **O2 U(iso) × 100 Å$^2$** | 0.81(3) | 0.93(3) | 0.95(3) | |
| **O3 (x,y,0)** | 0.07735(7) <br> 0.20678(7) | 0.07717(8) <br> 0.20693(7) | 0.07702(8) <br> 0.20691(7) | |
| **O3 U(iso) × 100 Å$^2$** | 1.18(2) | 1.28(3) | 1.36(2) | |
| **O4 (x,y, ½)** | 0.34481(8)<br>0.00716(7) | 0.34505(8)<br>0.00709(7) | 0.34518(8)<br>0.00717(7) | |
| **O4 U(iso) × 100 Å$^2$** | 1.06(3) | 1.15(3) | 1.25(3) | |
| **O5 (x,y, ½)** | 0.14079(7)<br>0.06873(8) | 0.14077(8)<br>0.06884(8) | 0.14102(7)<br>0.06901(8) | |
| **O5 U(iso) × 100 Å$^2$** | 0.91(2) | 1.02(3) | 1.14(2) | |

**Table S6:** Refinement details for $Ba_6InNb_9O_{30}$ at all temperatures in the non-polar space group P4/mbm with the B-site ion thermal parameters set as Uaniso.

| **Parameter** | **Temperature (K)** | | | |
|---|---|---|---|---|
| | **20** | **65** | **100** | **125** |
| **$\chi^2$** | 9.312 | 8.956 | 8.981 | 8.226 |
| **wRp (%)** | 5.45 | 5.34 | 5.36 | 5.17 |
| **Rp (%)** | 4.77 | 4.66 | 4.72 | 4.57 |
| **a (Å)** | 12.62644(3) | 12.62758(3) | 12.62929(3) | 12.63096(3) |
| **c (Å)** | 4.01148(22) | 4.01209(1) | 4.01288(2) | 4.01349(1) |
| **Cell Volume (Å$^3$)** | 639.538(3) | 639.751(3) | 640.051(3) | 640.317(3) |
| **Ba1 (0,0,0)** | | | | |
| **Ba1 U(iso) × 100 Å$^2$** | -0.24(4) | -0.20(4) | -0.18(4) | -0.15(4) |
| **Ba2 (x,y,0)** | 0.1709(1)<br>0.6709(1) | 0.1708(1)<br>0.6708(1) | 0.1708(1)<br>0.6708(1) | 0.1711(1)<br>0.6711(1) |
| **Ba2 U(iso) × 100 Å$^2$** | 1.58(4) | 1.68(4) | 1.71(4) | 1.77(4) |
| **Nb1/M$^{3+}$1, (0,½, ½)** | | | | |
| **Nb1/ M$^{3+}$1 U(aniso) × 100 Å$^2$**<br>**$U_{11}$**<br>**$U_{22}$**<br>**$U_{33}$**<br>**$U_{12}$**<br>**$U_{13}$**<br>**$U_{23}$** | <br>-0.01(5)<br>-0.01(5)<br>2.84(10)<br>-0.07(7)<br>0.0<br>0.0 | <br>-0.05(5)<br>-0.05(5)<br>3.01(10)<br>0.00(6)<br>0.0<br>0.0 | <br>0.02(5)<br>0.02(5)<br>2.8(1)<br>0.06(7)<br>0.0<br>0.0 | <br>0.05(5)<br>0.05(5)<br>2.8(1)<br>-0.04(6)<br>0.0<br>0.0 |
| **Nb2/ M$^{3+}$2 (x,y, ½)** | 0.07609(7)<br>0.21406(7) | 0.07603(7)<br>0.21417(6) | 0.07605(7)<br>0.21412(7) | 0.07606(7)<br>0.21412(6) |
| **Nb2/ M$^{3+}$2 U(aniso) × 100 Å$^2$**<br>**$U_{11}$**<br>**$U_{22}$**<br>**$U_{33}$**<br>**$U_{12}$**<br>**$U_{13}$**<br>**$U_{23}$** | <br>0.57(5)<br>0.14(4)<br>1.05(4)<br>0.28(4)<br>0.0<br>0.0 | <br>0.67(5)<br>0.17(4)<br>1.05(4)<br>0.28(4)<br>0.0<br>0.0 | <br>0.56(5)<br>0.22(4)<br>1.04(4)<br>0.29(4)<br>0.0<br>0.0 | <br>0.61(4)<br>0.22(4)<br>1.12(4)<br>0.27(4)<br>0.0<br>0.0 |
| **O1 (0,½,0)** | | | | |
| **O1 U(iso) × 100 Å$^2$** | 1.84(5) | 1.78(5) | 1.77(5) | 1.75(5) |
| **O2 (x,y, ½)** | 0.28096(8)<br>0.78096(8) | 0.28095(8)<br>0.78096(8) | 0.28094(8)<br>0.78094(8) | 0.28098(8)<br>0.78098(8) |
| **O2 U(iso) × 100 Å$^2$** | 0.71(4) | 0.72(4) | 0.74(4) | 0.73(4) |
| **O3 (x,y,0)** | 0.07673(9)<br>0.20694(9) | 0.07695(9)<br>0.20688(9) | 0.07699(9)<br>0.20689(9) | 0.07698(8)<br>0.20688(8) |
| **O3 U(iso) × 100 Å$^2$** | 1.37(3) | 1.38(3) | 1.31(3) | 1.34(3) |
| **O4 (x,y, ½)** | 0.34540(9)<br>0.00696(8) | 0.34529(8)<br>0.00688(7) | 0.34527(9)<br>0.00694(8) | 0.34524(8)<br>0.00690(7) |
| **O4 U(iso) × 100 Å$^2$** | 0.94(3) | 1.00(3) | 0.97(3) | 1.00(3) |
| **O5 (x,y, ½)** | 0.14006(8)<br>0.06806(9) | 0.14012(8)<br>0.06811(9) | 0.14003(8)<br>0.06807(9) | 0.14018(8)<br>0.06802(8) |
| **O5 U(iso) × 100 Å$^2$** | 0.84(3) | 0.84(3) | 0.80(3) | 0.85(3) |
| **Parameter** | **150** | **175** | **200** | **225** |
| **$\chi^2$** | 7.799 | 7.433 | 7.263 | 7.041 |
| **wRp (%)** | 4.96 | 4.93 | 4.78 | 4.71 |
| **Rp (%)** | 4.43 | 4.47 | 4.26 | 4.22 |
| **a (Å)** | 12.63278(2) | 12.63492(2) | 12.63720(2) | 12.63954(2) |
| **c (Å)** | 4.01418(1) | 4.01476(1) | 4.01536(1) | 4.01589(1) |
| **Cell Volume (Å$^3$)** | 640.611(3) | 640.922(3) | 641.249(3) | 641.570(3) |
| **Ba1 (0,0,0)** | | | | |
| **Ba1 U(iso) × 100 Å$^2$** | -0.10(4) | -0.05(4) | -0.07(4) | 0.01(4) |
| **Ba2 (x,y,0)** | 0.1716(1)<br>0.6716(1) | 0.1711(1)<br>0.6711(1) | 0.1711(1)<br>0.6711(1) | 0.1713(1)<br>0.6713(1) |
| **Ba2 U(iso) × 100 Å$^2$** | 1.83(4) | 1.96(4) | 1.99(4) | 2.01(4) |
| **Nb1/M$^{3+}$1, (0,½, ½)** | | | | |
| **Nb1/ M$^{3+}$1 U(aniso) × 100 Å$^2$**<br>**$U_{11}$**<br>**$U_{22}$**<br>**$U_{33}$**<br>**$U_{12}$**<br>**$U_{13}$**<br>**$U_{23}$** | <br>0.11(5)<br>0.11(5)<br>3.0(1)<br>0.04(6)<br>0.0<br>0.0 | <br>0.10(5)<br>0.10(5)<br>3.08(10)<br>0.01(6)<br>0.0<br>0.0 | <br>0.05(4)<br>0.05(4)<br>3.2(1)<br>0.05(6)<br>0.0<br>0.0 | <br>0.14(4)<br>0.14(4)<br>3.1(1)<br>0.07(6)<br>0.0<br>0.0 |
| **Nb2/ M$^{3+}$2 (x,y, ½)** | 0.07607(7)<br>0.21416(6) | 0.07615(7)<br>0.21411(6) | 0.07611(7)<br>0.21415(6) | 0.07597(7)<br>0.21428(6) |
| **Nb2/ M$^{3+}$2 U(aniso) × 100 Å$^2$**<br>**$U_{11}$**<br>**$U_{22}$**<br>**$U_{33}$**<br>**$U_{12}$**<br>**$U_{13}$**<br>**$U_{23}$** | <br>0.67(4)<br>0.30(4)<br>1.18(4)<br>0.24(4)<br>0.0<br>0.0 | <br>0.66(4)<br>0.32(4)<br>1.18(4)<br>0.30(3)<br>0.0<br>0.0 | <br>0.61(4)<br>0.30(4)<br>1.13(4)<br>0.21(3)<br>0.0<br>0.0 | <br>0.64(4)<br>0.31(4)<br>1.19(4)<br>0.21(3)<br>0.0<br>0.0 |
| **O1 (0,½,0)** | | | | |
| **O1 U(iso) × 100 Å$^2$** | 1.73(5) | 1.61(5) | 1.61(5) | 1.73(5) |
| **O2 (x,y, ½)** | 0.28096(7)<br>0.78096(7) | 0.28084(7)<br>0.78084(7) | 0.28091(7)<br>0.78091(7) | 0.28086(7)<br>0.78086(7) |
| **O2 U(iso) × 100 Å$^2$** | 0.79(3) | 0.83(3) | 0.79(3) | 0.87(3) |
| **O3 (x,y,0)** | 0.07695(8)<br>0.20698(8) | 0.07692(8)<br>0.20699(8) | 0.07684(8)<br>0.20701(8) | 0.07687(8)<br>0.20697(8) |
| **O3 U(iso) × 100 Å$^2$** | 1.36(3) | 1.35(3) | 1.32(3) | 1.28(3) |
| **O4 (x,y, ½)** | 0.34514(8)<br>0.00690(7) | 0.34525(8)<br>0.00689(7) | 0.34508(8)<br>0.00701(7) | 0.34519(8)<br>0.00685(7) |
| **O4 U(iso) × 100 Å$^2$** | 1.50(3) | 1.06(3) | 1.03(3) | 1.09(3) |
| **O5 (x,y, ½)** | 0.14016(8)<br>0.06827(8) | 0.14026(8)<br>0.06817(8) | 0.14026(8)<br>0.06828(8) | 0.14034(8)<br>0.06824(8) |
| **O5 U(iso) × 100 Å$^2$** | 0.90(3) | 0.89(3) | 0.90(3) | 0.97(3) |
| **Parameter** | **250** | **275** | **300** | **375** |
| **$\chi^2$** | 6.928 | 6.711 | 6.566 | 5.873 |
| **wRp (%)** | 4.67 | 4.62 | 4.58 | 4.37 |
| **Rp (%)** | 4.20 | 4.23 | 4.18 | 3.96 |
| **a (Å)** | 12.64198(2) | 12.64453(2) | 12.64716(2) | 12.65547(2) |
| **c (Å)** | 4.01632(1) | 4.01666(1) | 4.01690(1) | 4.01737(1) |
| **Cell Volume (Å$^3$)** | 641.888(3) | 642.200(3) | 642.506(3) | 643.425(2) |
| **Ba1 (0,0,0)** | | | | |
| **Ba1 U(iso) × 100 Å$^2$** | 0.03(4) | 0.06(4) | 0.13(4) | 0.34(4) |
| **Ba2 (x,y,0)** | 0.1713(1)<br>0.6713(1) | 0.1714(1)<br>0.6714(1) | 0.1714(1)<br>0.6714(1) | 0.1717(1)<br>0.6717(1) |
| **Ba2 U(iso) × 100 Å$^2$** | 2.11(4) | 2.19(4) | 2.25(4) | 2.51(4) |
| **Nb1/M$^{3+}$1, (0,½, ½)** | | | | |
| **Nb1/ M$^{3+}$1 U(aniso) × 100 Å$^2$**<br>**$U_{11}$**<br>**$U_{22}$**<br>**$U_{33}$**<br>**$U_{12}$**<br>**$U_{13}$**<br>**$U_{23}$** | <br>0.14(4)<br>0.14(4)<br>3.3(1)<br>-0.01(6)<br>0.0<br>0.0 | <br>0.23(4)<br>0.23(4)<br>3.2(1)<br>0.12(6)<br>0.0<br>0.0 | <br>0.22(5)<br>0.22(5)<br>3.4(1)<br>-0.04(6)<br>0.0<br>0.0 | <br>0.41(5)<br>0.41(5)<br>3.6(1)<br>0.01(6)<br>0.0<br>0.0 |
| **Nb2/ $M^{3+}$2 (x,y, ½)** | 0.07602(7)<br>0.21429(6) | 0.07591(7)<br>0.21441(6) | 0.07598(7)<br>0.21438(6) | 0.07575(7)<br>0.21459(6) |
| **Nb2/ $M^{3+}$2 U(aniso) × 100 Å$^2$**<br>**$U_{11}$**<br>**$U_{22}$**<br>**$U_{33}$**<br>**$U_{12}$**<br>**$U_{13}$**<br>**$U_{23}$** | <br>0.66(4)<br>0.41(4)<br>1.23(4)<br>0.19(3)<br>0.0<br>0.0 | <br>0.69(4)<br>0.45(4)<br>1.24(4)<br>0.21(3)<br>0.0<br>0.0 | <br>0.69(4)<br>0.44(4)<br>1.27(4)<br>0.18(3)<br>0.0<br>0.0 | <br>0.90(4)<br>0.49(4)<br>1.33(4)<br>0.25(3)<br>0.0<br>0.0 |
| **O1 (0,½,0)** | | | | |
| **O1 U(iso) × 100 Å$^2$** | 1.69(5) | 1.69(5) | 1.70(5) | 1.94(5) |
| **O2 (x,y, ½)** | 0.28095(7)<br>0.78095(7) | 0.28084(7)<br>0.78084(7) | 0.28096(7)<br>0.78096(7) | 0.28111(7)<br>0.78111(7) |
| **O2 U(iso) × 100 Å$^2$** | 0.87(3) | 0.96(3) | 0.97(3) | 1.20(3) |
| **O3 (x,y,0)** | 0.07694(8)<br>0.20706(8) | 0.07694(8)<br>0.20690(8) | 0.07688(8)<br>0.20696(8) | 0.07688(8)<br>0.20689(8) |
| **O3 U(iso) × 100 Å$^2$** | 1.37(3) | 1.39(3) | 1.41(3) | 1.57(3) |
| **O4 (x,y, ½)** | 0.34517(8)<br>0.00706(7) | 0.34514(8)<br>0.00699(7) | 0.34512(8)<br>0.00693(7) | 0.34535(8)<br>0.00690(7) |
| **O4 U(iso) × 100 Å$^2$** | 1.13(3) | 1.18(3) | 1.23(3) | 1.37(3) |
| **O5 (x,y, ½)** | 0.14047(8)<br>0.06834(8) | 0.14055(8)<br>0.06839(8) | 0.14062(8)<br>0.06841(8) | 0.14081(7)<br>0.06864(8) |
| **O5 U(iso) × 100 Å$^2$** | 0.97(3) | 1.04(3) | 1.10(3) | 1.24(3) |
| **Parameter** | **450** | | | |
| $\chi^2$ | 5.826 | | | |
| **wRp (%)** | 4.31 | | | |
| **Rp (%)** | 3.93 | | | |
| **a (Å)** | 12.66386(2) | | | |
| **c (Å)** | 4.01790(1) | | | |
| **Cell Volume (Å$^3$)** | 644.364(2) | | | |
| **Ba1 (0,0,0)** | | | | |
| **Ba1 U(iso) × 100 Å$^2$** | 0.33(4) | | | |
| **Ba2 (x,y,0)** | 0.1718(1)<br>0.6718(1) | | | |
| **Ba2 U(iso) × 100 Å$^2$** | 2.58(4) | | | |
| **Nb1/$M^{3+}$1, (0,½, ½)** | | | | |
| **Nb1/ $M^{3+}$1 U(aniso) × 100 Å$^2$**<br>**$U_{11}$**<br>**$U_{22}$**<br>**$U_{33}$**<br>**$U_{12}$**<br>**$U_{13}$**<br>**$U_{23}$** | <br>0.39(5)<br>0.39(5)<br>3.4(1)<br>-0.09(6)<br>0.0<br>0.0 | | | |
| **Nb2/ $M^{3+}$2 (x,y, ½)** | 0.07555(7)<br>0.21477(6) | | | |
| **Nb2/ $M^{3+}$2 U(aniso) × 100 Å$^2$**<br>**$U_{11}$**<br>**$U_{22}$**<br>**$U_{33}$**<br>**$U_{12}$**<br>**$U_{13}$**<br>**$U_{23}$** | <br>0.92(4)<br>0.48(4)<br>1.36(4)<br>0.21(3)<br>0.0<br>0.0 | | | |
| **O1 (0,½,0)** | | | | |
| **O1 U(iso) × 100 Å$^2$** | 2.11(5) | | | |
| **O2 (x,y, ½)** | 0.28139(7) | | | |
| **O2 U(iso) × 100 Å$^2$** | 1.23(3) | | | |
| **O3 (x,y,0)** | 0.07670(8) | | | |
| | 0.20696(8) | | | |
| **O3 U(iso) × 100 Å$^2$** | 1.66(3) | | | |
| **O4 (x,y, ½)** | 0.34528(8)<br>0.00694(7) | | | |
| **O4 U(iso) × 100 Å$^2$** | 1.48(3) | | | |
| **O5 (x,y, ½)** | 0.14090(8)<br>0.06874(8) | | | |
| **O5 U(iso) × 100 Å$^2$** | 1.36(3) | | | |

**Table S7:** Refinement details for $Ba_6GaNb_9O_{30}$ at all temperatures in the polar space group P4bm.

| **Parameter** | **Temperature (K)** | | | |
|---|---|---|---|---|
| | **10** | **40** | **72** | **100** |
| $\chi^2$ | 8.956 | 9.279 | 9.049 | 8.524 |
| **wRp (%)** | 5.03 | 5.11 | 5.04 | 4.89 |
| **Rp (%)** | 4.69 | 4.79 | 4.76 | 4.61 |
| **a (Å)** | 12.54968(3) | 12.55029(3) | 12.55181(3) | 12.55344 (3) |
| **c (Å)** | 3.97737(2) | 3.97761(2) | 3.97806(2) | 3.97862(2) |
| **Cell Volume (Å$^3$)** | 626.413(3) | 626.513(3) | 626.735(3) | 626.986(3) |
| **Ba1 (0,0,0)** | | | | |
| **Ba1 U(iso) × 100 Å$^2$** | -0.02(4) | 0.10(4) | 0.02(4) | 0.12(4) |
| **Ba2 (x,y,0)** | 0.1712(1)<br>0.6712(1) | 0.1713(1)<br>0.6713(1) | 0.1717(1)<br>0.6717(1) | 0.1714(1)<br>0.6714(1) |
| **Ba2 U(iso) × 100 Å$^2$** | 0.90(4) | 0.95(4) | 0.92(4) | 1.07(4) |
| **Nb1/M$^{3+}$1, (0,½,z)** | 0.518(2) | 0.519(2) | 0.521(2) | 0.519(2) |
| **Nb1/ M$^{3+}$1 U(iso) × 100 Å$^2$** | 0.80(5) | 0.80(5) | 0.74(5) | 0.80(5) |
| **Nb2/ M$^{3+}$2 (x,y,z)** | 0.07473(7)<br>0.21463(7)<br>0.510(1) | 0.07482(7)<br>0.21461(7)<br>0.512(1) | 0.07481(7)<br>0.21461(7)<br>0.511(1) | 0.07474(7)<br>0.21457(6)<br>0.513(1) |
| **Nb2/ M$^{3+}$2 U(iso) × 100 Å$^2$** | 0.42(3) | 0.42(3) | 0.44(3) | 0.48(3) |
| **O1 (0,½,z)** | -0.003(3) | -0.002(3) | -0.001(3) | -0.002(3) |
| **O1 U(iso) × 100 Å$^2$** | 1.53(5) | 1.53(5) | 1.47(5) | 1.41(5) |
| **O2 (x,y,z)** | 0.28240(8)<br>0.78240(8)<br>0.485(2) | 0.28234(8)<br>0.78234(8)<br>0.485(2) | 0.28225(8)<br>0.78225(8)<br>0.486(2) | 0.28226(8)<br>0.78226(8)<br>0.489(2) |
| **O2 U(iso) × 100 Å$^2$** | 0.26(4) | 0.27(4) | 0.21(4) | 0.32(4) |
| **O3 (x,y,z)** | 0.07678(9)<br>0.20767(8)<br>0.002(2) | 0.07667(9)<br>0.20792(8)<br>0.002(2) | 0.07689(9)<br>0.20768(8)<br>0.003(2) | 0.07671(8)<br>0.20761(8)<br>0.004(2) |
| **O3 U(iso) × 100 Å$^2$** | 0.80(3) | 0.83(3) | 0.78(3) | 0.81(3) |
| **O4 (x,y,z)** | 0.34362(8)<br>0.00736(7)<br>0.483(1) | 0.34386(8)<br>0.00729(8)<br>0.485(1) | 0.34365(8)<br>0.00728(8)<br>0.485(1) | 0.34366(8)<br>0.00729(7)<br>0.486(1) |
| **O4 U(iso) × 100 Å$^2$** | 0.52(3) | 0.53(3) | 0.58(3) | 0.52(3) |
| **O5 (x,y,z)** | 0.14097(8)<br>0.06942(8)<br>0.507(2) | 0.14097(8)<br>0.06931(8)<br>0.504(2) | 0.14116(8)<br>0.06934(8)<br>0.506(2) | 0.14108(8)<br>0.06936(8)<br>0.506(1) |
| **O5 U(iso) × 100 Å$^2$** | 0.26(3) | 0.35(3) | 0.33(3) | 0.38(2) |
| **Parameter** | **125** | **150** | **200** | **300** |
| $\chi^2$ | 8.901 | 8.800 | 7.994 | 13.23 |
| **wRp (%)** | 5.02 | 4.98 | 4.74 | 6.13 |
| **Rp (%)** | 4.75 | 4.68 | 4.55 | 6.00 |
| **a (Å)** | 12.55534(3) | 12.55750(3) | 12.56215(3) | 12.57234(3) |
| **c (Å)** | 3.97918(2) | 3.97975(2) | 3.98072(2) | 3.98183(2) |
| **Cell Volume (Å$^3$)** | 627.264(3) | 627.570(3) | 628.189(3) | 629.383(4) |
| **Ba1 (0,0,0)** | | | | |
| **Ba1 U(iso) × 100 Å$^2$** | 0.11(4) | 0.15(4) | 0.26(4) | 0.44(6) |
| **Ba2 (x,y,0)** | 0.1715(1) | 0.1716(1) | 0.1717(1) | 0.1721(1) |
| | 0.6715(1) | 0.6716(1) | 0.6717(1) | 0.6721(1) |
| **Ba2 U(iso) × 100 Å$^2$** | 1.15(4) | 1.18(4) | 1.33(4) | 1.70(6) |
| **Nb1/M$^{3+}$1, (0,½,z)** | 0.522(2) | 0.520(2) | 0.520(2) | 0.518(2) |
| **Nb1/ M$^{3+}$1 U(iso) × 100 Å$^2$** | 0.81(5) | 0.86(5) | 0.86(5) | 1.01(6) |
| **Nb2/ M$^{3+}$2 (x,y,z)** | 0.07455(7)<br>0.21455(7)<br>0.512(1) | 0.07467(7)<br>0.21456(7)<br>0.512(1) | 0.07458(7)<br>0.21471(6)<br>0.513(1) | 0.0745(1)<br>0.21485(9)<br>0.513(2) |
| **Nb2/ M$^{3+}$2 U(iso) × 100 Å$^2$** | 0.51 (3) | 0.56(3) | 0.60(3) | 0.68(3) |
| **O1 (0,½,z)** | -0.001(3) | -0.003(3) | -0.004(3) | -0.002(4) |
| **O1 U(iso) × 100 Å$^2$** | 1.41(5) | 1.43(5) | 1.38(5) | 1.46(7) |
| **O2 (x,y,z)** | 0.28230(8)<br>0.78231(8)<br>0.489(2) | 0.28239(8)<br>0.78239(8)<br>0.489(2) | 0.28224(8)<br>0.78224(8)<br>0.490(2) | 0.2825(1)<br>0.7825(1)<br>0.487(2) |
| **O2 U(iso) × 100 Å$^2$** | 0.36(4) | 0.37(4) | 0.48(4) | 0.54(5) |
| **O3 (x,y,z)** | 0.07668(9)<br>0.20776(8)<br>0.004(2) | 0.07681(9)<br>0.20774(8)<br>0.005(2) | 0.07690(8)<br>0.20761(8)<br>0.003(2) | 0.0768(1)<br>0.2078(1)<br>0.002(3) |
| **O3 U(iso) × 100 Å$^2$** | 0.87(3) | 0.81(3) | 0.80(3) | 0.91(3) |
| **O4 (x,y,z)** | 0.34376(8)<br>0.00746(8)<br>0.487(1) | 0.34362(8)<br>0.00737(8)<br>0.486(1) | 0.34374(8)<br>0.00734(7)<br>0.487(2) | 0.3435(1)<br>0.0072(1)<br>0.488(2) |
| **O4 U(iso) × 100 Å$^2$** | 0.64(3) | 0.63(3) | 0.67(3) | 0.92(4) |
| **O5 (x,y,z)** | 0.14128(8)<br>0.06925(8)<br>0.505(2) | 0.14116(8)<br>0.06933(8)<br>0.507(2) | 0.14133(8)<br>0.06937(8)<br>0.504(2) | 0.1416(1)<br>0.0696(1)<br>0.501(2) |
| **O5 U(iso) × 100 Å$^2$** | 0.40(3) | 0.43(3) | 0.48(2) | 0.68(3) |

**Table S8:** Refinement details for $Ba_6ScNb_9O_{30}$ at all temperatures in the polar space group P4bm.

| **Parameter** | **Temperature (K)** | | | |
|---|---|---|---|---|
| | **8** | **100** | **150** | **180** |
| **$\chi^2$** | 11.37 | 10.76 | 10.13 | 9.506 |
| **wRp (%)** | 5.54 | 5.35 | 5.19 | 5.00 |
| **Rp (%)** | 5.01 | 4.88 | 4.79 | 4.68 |
| **a (Å)** | 12.60942(3) | 12.61243(3) | 12.61594(3) | 12.61850(3) |
| **c (Å)** | 4.00266(2) | 4.00406(1) | 4.00530(1) | 4.00598(1) |
| **Cell Volume (Å$^3$)** | 636.413(3) | 636.939(3) | 637.491(3) | 637.858(3) |
| **Ba1 (0,0,0)** | | | | |
| **Ba1 U(iso) × 100 Å$^2$** | -0.16(4) | -0.14(4) | -0.06(4) | 0.03(4) |
| **Ba2 (x,y,0)** | 0.1708(1)<br>0.6708(1) | 0.1710(1)<br>0.6710(1) | 0.1711(1)<br>0.6711(1) | 0.1710(1)<br>0.6710(1) |
| **Ba2 U(iso) × 100 Å$^2$** | 1.15(4) | 1.24(4) | 1.41(4) | 1.55(4) |
| **Nb1/M$^{3+}$1, (0,½, z)** | 0.533(1) | 0.532(1) | 0.534(1) | 0.534(1) |
| **Nb1/ M$^{3+}$1 U(iso) × 100 Å$^2$** | 0.71(4) | 0.79(4) | 0.82(4) | 0.88(4) |
| **Nb2/ M$^{3+}$2 (x,y,z)** | 0.07552(6)<br>0.21434(6)<br>0.508(1) | 0.07551(6)<br>0.21429(6)<br>0.506(1) | 0.07551(6)<br>0.21441(6)<br>0.507(1) | 0.07547(6)<br>0.21437(6)<br>0.506(1) |
| **Nb2/ M$^{3+}$2 U(iso) × 100 Å$^2$** | 0.40(2) | 0.45(2) | 0.48(2) | 0.55(2) |
| **O1 (0,½,z)** | -0.001(2) | -0.001(2) | -0.001(2) | 0.0002(2) |
| **O1 U(iso) × 100 Å$^2$** | 1.75(5) | 1.56(5) | 1.55(5) | 1.52(5) |
| **O2 (x,y, z)** | 0.28134(7)<br>0.78134(7)<br>0.483(2) | 0.28119(7)<br>0.78119(7)<br>0.482(1) | 0.28128(7)<br>0.78128(7)<br>0.485(2) | 0.28122(7)<br>0.78122(7)<br>0.485(1) |
| **O2 U(iso) × 100 Å$^2$** | 0.29(4) | 0.28(3) | 0.39(3) | 0.37(3) |
| **O3 (x,y,z)** | 0.07711(8)<br>0.20676(8)<br>-0.001(2) | 0.07725(8)<br>0.20669(8)<br>-0.003(2) | 0.07736(8)<br>0.20677(8)<br>-0.001(2) | 0.07726(8)<br>0.20689(7)<br>-0.001(2) |
| **O3 U(iso) × 100 Å$^2$** | 1.07(3) | 1.01(3) | 0.99(3) | 1.02(3) |
| **O4 (x,y,z)** | 0.34453(8)<br>0.00750(7)<br>0.494(1) | 0.34464(8)<br>0.00737(7)<br>0.499(1) | 0.34464(8)<br>0.00743(7)<br>0.495(1) | 0.34452(8)<br>0.00734(7)<br>0.499(1) |
| **O4 U(iso) × 100 Å$^2$** | 0.91(3) | 0.90(3) | 0.88(3) | 0.95(3) |
| **O5 (x,y,z)** | 0.14036(8)<br>0.06845(8)<br>0.491(1) | 0.14042(8)<br>0.06842(8)<br>0.489(1) | 0.14044(7)<br>0.06853(8)<br>0.491(1) | 0.14058(7)<br>0.06863(8)<br>0.491(1) |
| **O5 U(iso) × 100 Å$^2$** | 0.56(3) | 0.52(3) | 0.61(3) | 0.62(2) |
| **Parameter** | **200** | **220** | **240** | **260** |
| **$\chi^2$** | 9.776 | 9.377 | 9.136 | 8.840 |
| **wRp (%)** | 5.07 | 4.96 | 4.90 | 4.83 |
| **Rp (%)** | 4.67 | 4.63 | 4.53 | 4.50 |
| **a (Å)** | 12.62028(3) | 12.62215(3) | 12.62406(3) | 12.62596(3) |
| **c (Å)** | 4.00638(1) | 4.00671(1) | 4.00697(1) | 4.00721(1) |
| **Cell Volume (Å$^3$)** | 638.102(3) | 638.343(3) | 638.578(3) | 638.808(3) |
| **Ba1 (0,0,0)** | | | | |
| **Ba1 U(iso) × 100 Å$^2$** | 0.02(4) | 0.01(4) | 0.11(4) | 0.21(4) |
| **Ba2 (x,y,0)** | 0.1712(1)<br>0.6711(1) | 0.1712(1)<br>0.6712(1) | 0.1712(1)<br>0.6712(1) | 0.1712(1)<br>0.6712(1) |
| **Ba2 U(iso) × 100 Å$^2$** | 1.43(4) | 1.55(4) | 1.66(4) | 1.87(4) |
| **Nb1/M$^{3+}$1, (0,½,z)** | 0.534(1) | 0.535(1) | 0.534(1) | 0.535(1) |
| **Nb1/ M$^{3+}$1 U(iso) × 100 Å$^2$** | 0.82(4) | 0.88(4) | 0.89(4) | 1.08(4) |
| **Nb2/ M$^{3+}$2 (x,y,z)** | 0.07548(6)<br>0.21436(6)<br>0.504(1) | 0.07540(6)<br>0.21442(6)<br>0.505(1) | 0.07543(6)<br>0.21446(5)<br>0.506(1) | 0.07536(6)<br>0.21474(6)<br>0.512(1) |
| **Nb2/ M$^{3+}$2 U(iso) × 100 Å$^2$** | 0.47(2) | 0.53(2) | 0.53(2) | 0.66(2) |
| **O1 (0,½,z)** | -0.001(2) | 0.001(2) | 0.000(2) | -0.002(2) |
| **O1 U(iso) × 100 Å$^2$** | 1.50(5) | 1.47(5) | 1.53(5) | 1.57(5) |
| **O2 (x,y,z)** | 0.28131(7)<br>0.78131(7)<br>0.487(2) | 0.28115(7)<br>0.78115(7)<br>0.486(2) | 0.28130(7)<br>0.78130(7)<br>0.487(2) | 0.28106(7)<br>0.78106(7)<br>0.498(2) |
| **O2 U(iso) × 100 Å$^2$** | 0.40(3) | 0.48(3) | 0.50(3) | 0.74(3) |
| **O3 (x,y,z)** | 0.07736(8)<br>0.20674(8)<br>-0.003(2) | 0.07732(8)<br>0.20681(7)<br>-0.001(2) | 0.07741(7)<br>0.20680(7)<br>-0.001(2) | 0.07726(8)<br>0.20670(8)<br>0.004(2) |
| **O3 U(iso) × 100 Å$^2$** | 0.91(3) | 0.98(3) | 0.99(3) | 1.14(3) |
| **O4 (x,y,z)** | 0.34469(8)<br>0.00727(7)<br>0.498(1) | 0.34463(8)<br>0.00739(7)<br>0.499(1) | 0.34462(8)<br>0.00731(7)<br>0.497(1) | 0.34458(8)<br>0.00725(7)<br>0.492(1) |
| **O4 U(iso) × 100 Å$^2$** | 0.85(3) | 0.95(3) | 0.97(3) | 1.05(3) |
| **O5 (x,y,z)** | 0.14043(7)<br>0.06858(8)<br>0.488(1) | 0.14063(7)<br>0.06863(8)<br>0.491(1) | 0.14064(7)<br>0.06869(8)<br>0.490(1) | 0.14090(8)<br>0.06887(8)<br>0.493(1) |
| **O5 U(iso) × 100 Å$^2$** | 0.55(3) | 0.64(2) | 0.66(2) | 0.78(3) |
| **Parameter** | **300** | **375** | **450** | |
| **$\chi^2$** | 8.065 | 8.112 | 7.494 | |
| **wRp (%)** | 4.63 | 4.62 | 4.44 | |
| **Rp (%)** | 4.30 | 4.21 | 4.03 | |
| **a (Å)** | 12.63005(2) | 12.63813(2) | 12.64615(2) | |
| **c (Å)** | 4.00749(1) | 4.00800(1) | 4.00858(1) | |
| **Cell Volume (Å$^3$)** | 639.268(3) | 640.168(3) | 641.072(3) | |
| **Ba1 (0,0,0)** | | | | |
| **Ba1 U(iso) × 100 Å$^2$** | 0.20(4) | 0.31(4) | 0.45(4) | |
| **Ba2 (x,y,0)** | 0.1713(1)<br>0.6713(1) | 0.1715(1)<br>0.6715(1) | 0.1716(1)<br>0.6716(1) | |
| **Ba2 U(iso) × 100 Å$^2$** | 1.95(4) | 1.97(4) | 2.25(4) | |
| **Nb1/M$^{3+}$1, (0,½,z)** | 0.534(1) | 0.531(1) | 0.529(1) | |
| **Nb1/ $M^{3+}$1 U(iso) × 100 Å$^2$** | 1.14(4) | 1.21(4) | 1.27(4) | |
| **Nb2/ $M^{3+}$2 (x,y,z)** | 0.07539(6)<br>0.21475(6)<br>0.509(1) | 0.07535(6)<br>0.21484(5)<br>0.508(1) | 0.07527(6)<br>0.21505(5)<br>0.508(1) | |
| **Nb2/ $M^{3+}$2 U(iso) × 100 Å$^2$** | 0.68(2) | 0.64(2) | 0.71(2) | |
| **O1 (0,½,z)** | 0.002(2) | -0.001(2) | 0.001(2) | |
| **O1 U(iso) × 100 Å$^2$** | 1.70(5) | 1.93(5) | 2.06(5) | |
| **O2 (x,y,z)** | 0.28120(7)<br>0.78120(7)<br>0.486(2) | 0.28152(7)<br>0.78153(7)<br>0.488(2) | 0.28159(7)<br>0.78159(7)<br>0.486(2) | |
| **O2 U(iso) × 100 Å$^2$** | 0.68(4) | 0.74(4) | 0.81(4) | |
| **O3 (x,y,z)** | 0.07734(8)<br>0.20685(8)<br>0.002(2) | 0.07729(8)<br>0.20700(7)<br>0.001(2) | 0.07705(8)<br>0.20704(8)<br>0.000(2) | |
| **O3 U(iso) × 100 Å$^2$** | 1.18(3) | 1.21(3) | 1.37(3) | |
| **O4 (x,y,z)** | 0.34465(8)<br>0.00732(7)<br>0.497(1) | 0.34489(8)<br>0.00731(7)<br>0.494(1) | 0.34495(8)<br>0.00734(7)<br>0.489(1) | |
| **O4 U(iso) × 100 Å$^2$** | 1.13(3) | 1.19(3) | 1.32(3) | |
| **O5 (x,y,z)** | 0.14085(7)<br>0.06891(8)<br>0.495(1) | 0.14074(7)<br>0.06897(8)<br>0.493(2) | 0.14097(7)<br>0.06914(8)<br>0.494(2) | |
| **O5 U(iso) × 100 Å$^2$** | 0.85(3) | 0.91(3) | 1.07(3) | |

**Table S9:** Refinement details for $Ba_6InNb_9O_{30}$ at all temperatures in the polar space group P4bm.

| **Parameter** | **Temperature (K)** | | | |
|---|---|---|---|---|
| | **20** | **65** | **100** | **125** |
| **$\chi^2$** | 9.290 | 8.983 | 8.986 | 8.231 |
| **wRp (%)** | 5.45 | 5.35 | 5.36 | 5.17 |
| **Rp (%)** | 4.57 | 4.68 | 4.72 | 4.58 |
| **a (Å)** | 12.62638(3) | 12.62752(3) | 12.62925(3) | 12.63091(3) |
| **c (Å)** | 4.01149(2) | 4.01209(1) | 4.01288(2) | 4.01349(1) |
| **Cell Volume (Å$^3$)** | 639.533(3) | 639.746(3) | 640.046(3) | 640.312(3) |
| **Ba1 (0,0,0)** | | | | |
| **Ba1 U(iso) × 100 Å$^2$** | -0.10(4) | -0.07(4) | -0.04(4) | -0.02(4) |
| **Ba2 (x,y,0)** | 0.1709(1)<br>0.6709(1) | 0.1708(1)<br>0.6708(1) | 0.1708(1)<br>0.6708(1) | 0.1710(1)<br>0.6710(1) |
| **Ba2 U(iso) × 100 Å$^2$** | 1.64(4) | 1.70(4) | 1.72(4) | 1.80(4) |
| **Nb1/$M^{3+}$1, (0,½, z)** | 0.532(1) | 0.534(1) | 0.532(1) | 0.532(1) |
| **Nb1/ $M^{3+}$1 U(iso) × 100 Å$^2$** | 0.32(4) | 0.28(4) | 0.31(4) | 0.36(4) |
| **Nb2/ $M^{3+}$2 (x,y,z)** | 0.07599(7)<br>0.21431(7)<br>0.512(1) | 0.07596(7)<br>0.21439(7)<br>0.512(1) | 0.07596(7)<br>0.21433(7)<br>0.510(1) | 0.07597(7)<br>0.21436(6)<br>0.512(1) |
| **Nb2/ $M^{3+}$2 U(iso) × 100 Å$^2$** | 0.48(3) | 0.51(3) | 0.51(3) | 0.54(3) |
| **O1 (0,½,z)** | -0.006(2) | -0.004(2) | -0.004(2) | -0.005(2) |
| **O1 U(iso) × 100 Å$^2$** | 2.10(6) | 2.03(6) | 1.99(6) | 1.98(5) |
| **O2 (x,y, z)** | 0.28083(8)<br>0.78084(8)<br>0.502(2) | 0.28084(8)<br>0.78085(8)<br>0.497(2) | 0.28081(8)<br>0.78082(8)<br>0.496(2) | 0.28086(7)<br>0.78087(7)<br>0.500(2) |
| **O2 U(iso) × 100 Å$^2$** | 0.69(4) | 0.68(4) | 0.70(4) | 0.71(4) |
| **O3 (x,y,z)** | 0.07703(9)<br>0.20692(9)<br>0.000(1) | 0.07717(9)<br>0.20688(9)<br>0.002(2) | 0.07719(9)<br>0.20686(9)<br>-0.001(2) | 0.07724(8)<br>0.20686(8)<br>0.000(2) |
| **O3 U(iso) × 100 Å$^2$** | 1.35(3) | 1.37(3) | 1.30(3) | 1.32(3) |
| **O4 (x,y,z)** | 0.34523(8)<br>0.00717(8)<br>0.484(1) | 0.34510(8)<br>0.00709(8)<br>0.487(1) | 0.34512(8)<br>0.00716(8)<br>0.487(1) | 0.34508(8)<br>0.00710(7)<br>0.486(1) |
| **O4 U(iso) × 100 Å$^2$** | 0.90(3) | 1.00(3) | 0.98(3) | 0.99(3) |
| **O5 (x,y,z)** | 0.14010(8)<br>0.06815(9)<br>0.488(1) | 0.14015(8)<br>0.06821(9)<br>0.4896(12) | 0.14009(8)<br>0.06816(9)<br>0.488(1) | 0.14023(8)<br>0.06811(8)<br>0.488(1) |
| **O5 U(iso) × 100 Å$^2$** | 0.76(3) | 0.77(3) | 0.73(3) | 0.77(3) |
| **Parameter** | **150** | **175** | **200** | **225** |
| **$\chi^2$** | 7.841 | 7.454 | 7.313 | 7.114 |
| **wRp (%)** | 4.97 | 4.94 | 4.79 | 4.74 |
| **Rp (%)** | 4.44 | 4.46 | 4.26 | 4.24 |
| **a (Å)** | 12.63273(2) | 12.63488(2) | 12.63716(2) | 12.63949(2) |
| **c (Å)** | 4.01418(1) | 4.01476(1) | 4.01536(1) | 4.01589(1) |
| **Cell Volume (Å$^3$)** | 640.606(3) | 640.916(3) | 641.244(3) | 641.566(3) |
| **Ba1 (0,0,0)** | | | | |
| **Ba1 U(iso) × 100 Å$^2$** | 0.03(4) | 0.08(4) | 0.06(4) | 0.13(4) |
| **Ba2 (x,y,0)** | 0.1711(1)<br>0.6711(1) | 0.1710(1)<br>0.6710(1) | 0.1710(1)<br>0.6710(1) | 0.1713(1)<br>0.6713(1) |
| **Ba2 U(iso) × 100 Å$^2$** | 1.85(4) | 1.98(4) | 1.95(4) | 1.98(4) |
| **Nb1/M$^{3+}$1, (0,½,z)** | 0.533(1) | 0.536(1) | 0.535(1) | 0.534(1) |
| **Nb1/ M$^{3+}$1 U(iso) × 100 Å$^2$** | 0.41(4) | 0.36(4) | 0.34(4) | 0.40(4) |
| **Nb2/ M$^{3+}$2 (x,y,z)** | 0.07600(7)<br>0.21436(6)<br>0.512(1) | 0.07606(7)<br>0.21432(6)<br>0.513(3) | 0.07603(7)<br>0.21433(6)<br>0.510(1) | 0.07589(7)<br>0.21447(6)<br>0.511(1) |
| **Nb2/ M$^{3+}$2 U(iso) × 100 Å$^2$** | 0.61(3) | 0.60(3) | 0.59(3) | 0.59(3) |
| **O1 (0,½,z)** | -0.006(2) | -0.004(2) | -0.004(2) | -0.006(2) |
| **O1 U(iso) × 100 Å$^2$** | 1.96(5) | 1.84(5) | 1.80(5) | 1.89(5) |
| **O2 (x,y,z)** | 0.28086(7)<br>0.78087(7)<br>0.498(2) | 0.28073(7)<br>0.78074(7)<br>0.495(2) | 0.28082(7)<br>0.78082(7)<br>0.493(2) | 0.28077(7)<br>0.78078(7)<br>0.502(2) |
| **O2 U(iso) × 100 Å$^2$** | 0.76(3) | 0.77(4) | 0.73(3) | 0.81(3) |
| **O3 (x,y,z)** | 0.07717(8)<br>0.20697(8)<br>0.001(2) | 0.07710(8)<br>0.20696(8)<br>0.002(2) | 0.07701(8)<br>0.20698(8)<br>-0.001(2) | 0.07709(8)<br>0.20696(8)<br>-0.001(2) |
| **O3 U(iso) × 100 Å$^2$** | 1.34(3) | 1.34(3) | 1.30(3) | 1.23(3) |
| **O4 (x,y,z)** | 0.34498(8)<br>0.00710(7)<br>0.487(1) | 0.34508(8)<br>0.00711(7)<br>0.489(1) | 0.34494(8)<br>0.00721(7)<br>0.490(1) | 0.34504(8)<br>0.00701(7)<br>0.487(1) |
| **O4 U(iso) × 100 Å$^2$** | 1.04(3) | 1.07(3) | 1.06(3) | 1.05(3) |
| **O5 (x,y,z)** | 0.14021(8)<br>0.06836(8)<br>0.490(1) | 0.14036(8)<br>0.06826(8)<br>0.492(1) | 0.14031(7)<br>0.06839(8)<br>0.489(1) | 0.14039(7)<br>0.06834(8)<br>0.488(1) |
| **O5 U(iso) × 100 Å$^2$** | 0.84(3) | 0.84(3) | 0.82(3) | 0.85(3) |
| **Parameter** | **250** | **275** | **300** | **375** |
| **$\chi^2$** | 7.031 | 6.813 | 6.662 | 6.027 |
| **wRp (%)** | 4.71 | 4.66 | 4.61 | 4.43 |
| **Rp (%)** | 4.22 | 4.26 | 4.19 | 3.98 |
| **a (Å)** | 12.64194(2) | 12.64448(2) | 12.64711(2) | 12.65541(2) |
| **c (Å)** | 4.01632(1) | 4.01666(1) | 4.01690(1) | 4.01737(1) |
| **Cell Volume (Å$^3$)** | 641.883(3) | 642.195(3) | 642.501(3) | 643.419(2) |
| **Ba1 (0,0,0)** | | | | |
| **Ba1 U(iso) × 100 Å$^2$** | 0.13(4) | 0.16(4) | 0.23(4) | 0.43(4) |
| **Ba2 (x,y,0)** | 0.1712(1)<br>0.6712(1) | 0.1713(1)<br>0.6713(1) | 0.1713(1)<br>0.6713(1) | 0.1716(1)<br>0.6716(1) |
| **Ba2 U(iso) × 100 Å$^2$** | 2.03(4) | 2.12(4) | 2.17(4) | 2.42(4) |
| **Nb1/M$^{3+}$1, (0,½,z)** | 0.534(1) | 0.535(1) | 0.536(1) | 0.536(1) |
| **Nb1/ M$^{3+}$1 U(iso) × 100 Å$^2$** | 0.41(4) | 0.46(4) | 0.46(4) | 0.65(4) |
| **Nb2/ M$^{3+}$2 (x,y,z)** | 0.07593(7)<br>0.21446(6) | 0.07586(7)<br>0.21452(6) | 0.07593(7)<br>0.21450(6) | 0.07570(7)<br>0.21470(6) |
| | 0.508(1) | 0.509(1) | 0.508(1) | 0.508(1) |
| **Nb2/ $M^{3+}$2 U(iso) × 100 Å$^2$** | 0.65(3) | 0.69(3) | 0.70(3) | 0.79(2) |
| **O1 (0,½,z)** | -0.003(2) | -0.003(2) | -0.002(2) | -0.001(2) |
| **O1 U(iso) × 100 Å$^2$** | 1.82(5) | 1.83(5) | 1.85(5) | 2.07(5) |
| **O2 (x,y,z)** | 0.28089(7)<br>0.78089(7)<br>0.489(2) | 0.28080(7)<br>0.78080(7)<br>0.490(2) | 0.28093(7)<br>0.78094(7)<br>0.485(2) | 0.28105(7)<br>0.78106(7)<br>0.489(2) |
| **O2 U(iso) × 100 Å$^2$** | 0.75(4) | 0.83(4) | 0.79(4) | 1.05(4) |
| **O3 (x,y,z)** | 0.07708(8)<br>0.20703(8)<br>-0.003(2) | 0.07705(8)<br>0.20688(8)<br>-0.002(2) | 0.07696(8)<br>0.20697(8)<br>-0.002(2) | 0.07695(8)<br>0.20693(8)<br>0.000(2) |
| **O3 U(iso) × 100 Å$^2$** | 1.31(3) | 1.34(3) | 1.35(3) | 1.52(3) |
| **O4 (x,y,z)** | 0.34504(8)<br>0.00726(7)<br>0.491(1) | 0.34501(8)<br>0.00719(7)<br>0.492(1) | 0.34497(8)<br>0.00715(7)<br>0.494(1) | 0.34518(8)<br>0.00710(7)<br>0.495(1) |
| **O4 U(iso) × 100 Å$^2$** | 1.15(3) | 1.20(3) | 1.27(3) | 1.41(3) |
| **O5 (x,y,z)** | 0.14054(7)<br>0.06846(8)<br>0.489(1) | 0.14061(7)<br>0.06848(8)<br>0.491(1) | 0.14069(8)<br>0.06852(8)<br>0.491(1) | 0.14084(7)<br>0.06877(8)<br>0.491(1) |
| **O5 U(iso) × 100 Å$^2$** | 0.86(3) | 0.94(3) | 1.00(3) | 1.11(3) |
| **Parameter** | **450** | | | |
| **$\chi^2$** | 5.938 | | | |
| **wRp (%)** | 4.35 | | | |
| **Rp (%)** | 3.94 | | | |
| **a (Å)** | 12.66380(2) | | | |
| **c (Å)** | 4.01789(1) | | | |
| **Cell Volume (Å$^3$)** | 644.357(2) | | | |
| **Ba1 (0,0,0)** | | | | |
| **Ba1 U(iso) × 100 Å$^2$** | 0.45(4) | | | |
| **Ba2 (x,y,0)** | 0.1717(1)<br>0.6717(1) | | | |
| **Ba2 U(iso) × 100 Å$^2$** | 2.56(4) | | | |
| **Nb1/$M^{3+}$1, (0,½,z)** | 0.535(1) | | | |
| **Nb1/ $M^{3+}$1 U(iso) × 100 Å$^2$** | 0.70(4) | | | |
| **Nb2/ $M^{3+}$2 (x,y,z)** | 0.07553(7)<br>0.21492(6)<br>0.511(1) | | | |
| **Nb2/ $M^{3+}$2 U(iso) × 100 Å$^2$** | 0.82(3) | | | |
| **O1 (0,½,z)** | -0.001(2) | | | |
| **O1 U(iso) × 100 Å$^2$** | 2.33(5) | | | |
| **O2 (x,y,z)** | 0.28137(7)<br>0.78137(7)<br>0.488(2) | | | |
| **O2 U(iso) × 100 Å$^2$** | 1.12(4) | | | |
| **O3 (x,y,z)** | 0.07678(8)<br>0.20700(8)<br>0.000(2) | | | |
| **O3 U(iso) × 100 Å$^2$** | 1.65(3) | | | |
| **O4 (x,y,z)** | 0.34511(8)<br>0.00715(7)<br>0.493(1) | | | |
| **O4 U(iso) × 100 Å$^2$** | 1.54(3) | | | |
| **O5 (x,y,z)** | 0.14089(8)<br>0.06888(8)<br>0.492(1) | | | |
| **O5 U(iso) × 100 Å$^2$** | 1.27(3) | | | |

**Calculation of octahedral rotation in TTB materials**

There appears to be no precedent for calculating octahedral rotation in TTB materials. In order to address the rotation of B2 octahedra we can simplify the structure somewhat: since the B1 site is crystallographically fixed, we can treat the B2 octahedra separately as they essentially form the perovskite building block running along the c-axis of the TTB cell. It should therefore be possible be able to describe the tilting of the octahedral units using the ideas laid out for perovskites.

The TTB structures here have space groups P4/mbm and P4bm both of which have Glazer notation $a^0a^0c^+$. This means that the tilts are in-phase in the $[001]_p$ (c-axis). The octahedral tilt in tetragonal perovskites may be calculated from the lattice parameters or from the atomic positions of the oxygen ions. Both methods are problematic with the TTB structure and are summarised below.

(a) Octahedral tilts from the lattice parameters.

The octahedral tilts for tetragonal perovskites ($a^0a^0c^+$) can be related to the lattice parameter (a) by equation 1.

$$a = 8^{1/2} d \cos\varphi \qquad -(1)$$

Where a is the lattice parameter, d is the epitaxial M-O bond length and φ is the octahedral tilt angle.

This approach gives rise to two problems with respect to TTBs. Firstly the lattice parameters do not relate directly to the perovskite unit. Secondly in the perovskite the epitaxial M-O bond lengths is described by one oxygen crystallographic position with equal bond lengths; in the TTB structure, however, the epitaxial oxygen atoms are described by three crystallographic positions with four different M-O bond lengths.

As a further simplification it is possible to use an average bond length and the lattice parameter can be determined from the TTB structure as the diagonal of the perovskite unit (see figure S1).

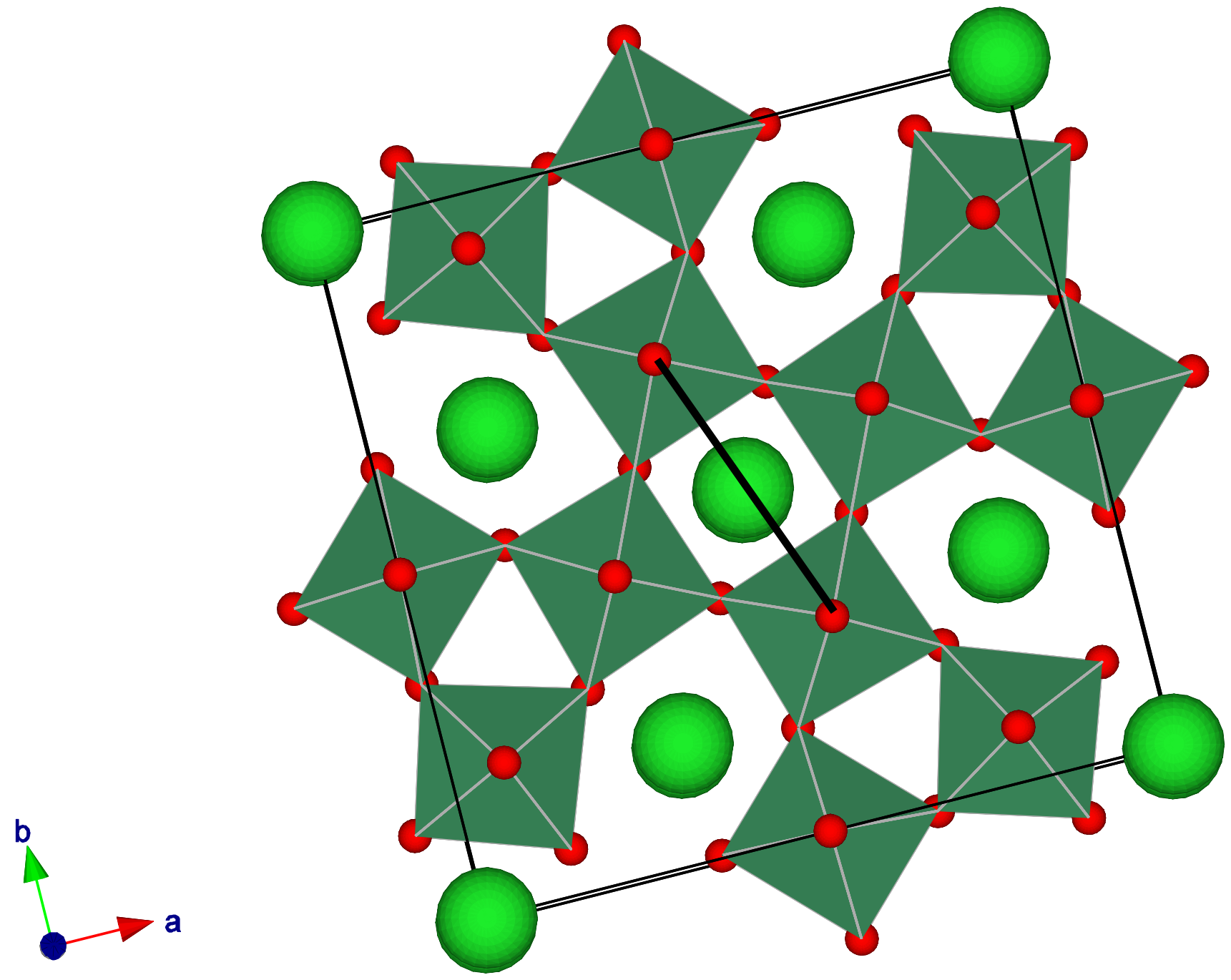

**Figure S1:** TTB structure showing the perovskite lattice parameter marked with a line

Calculations done using this method give tilt angles comparable to those already in the literature for tetragonal perovskites. The tilt angles as a function of temperature show that with increasing temperature the tilt angle increases for all compositions (as shown in figure S2).

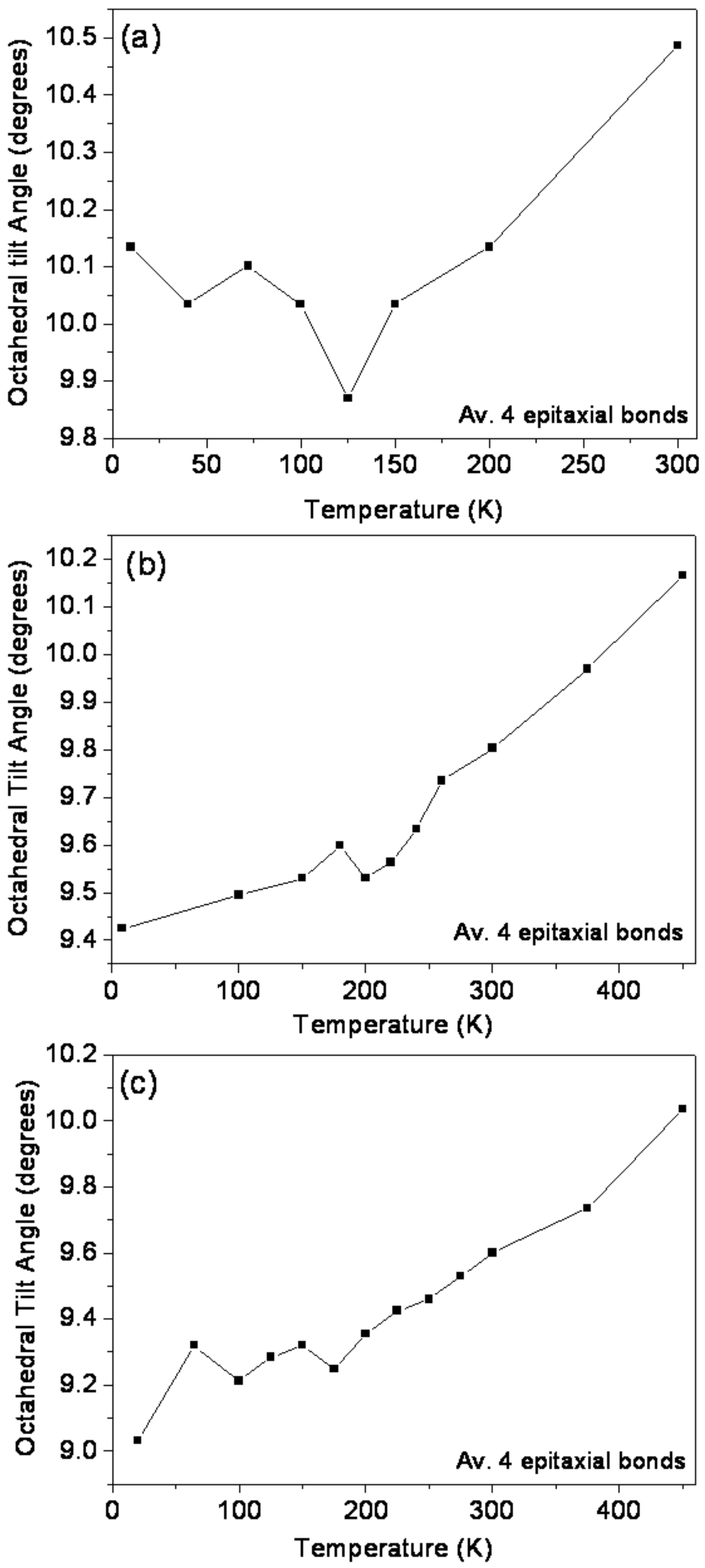

**Figure S2:** Variation of the B2 octahedral tilt angle with temperature as calculated using the lattice parameters for (a) $Ba_6GaNb_9O_{30}$, (b) $Ba_6ScNb_9O_{30}$ and (c) $Ba_6InNb_9O_{30}$.

(b) Octahedral tilts from the atomic positions.

It is also possible to calculate the octahedral rotation from the atomic positions of the oxygen atoms. This is fairly simple in perovskites as this is calculated directly from the relationship between the atomic positions of the O2 ion in comparison with the ideal O2 position. However in the TTB structure whilst the O2 position describes one of the epitaxial M-O bonds the other 3 are described by O4 and O5 and oxygen atomic positions. Ideally the octahedral tilt should be calculated taking into account all the epitaxial oxygen positions and their relationships to the ideal positions. This is rather complex and a simplification is to calculate tilts from the O2 atomic positions which can be described by the atomic position ¼ + u, ¾ + u, ½ where u can be used to calculate the octahedral tilt from equation 2.

$$\varphi = \tan^{-1}(4u) \qquad -(2)$$

The results are shown in figure S3. The tilt increases with increasing temperature for all compositions, consistent with those calculated from lattice parameters. However, the tilt magnitudes are smaller than those calculated using the lattice parameters, which are more consistent with magnitudes observed in perovskite oxides.

While the tilt magnitudes may not be accurate the results clearly indicate that the tilt increases with temperature. There is no obvious link to the electrical properties although the tilt angle does appear to increase more sharply above $T_{c/a}$. It should be noted that none of these calculations take into account that the $M^{3+}$ is randomly distributed across the B1 and B2 sites.

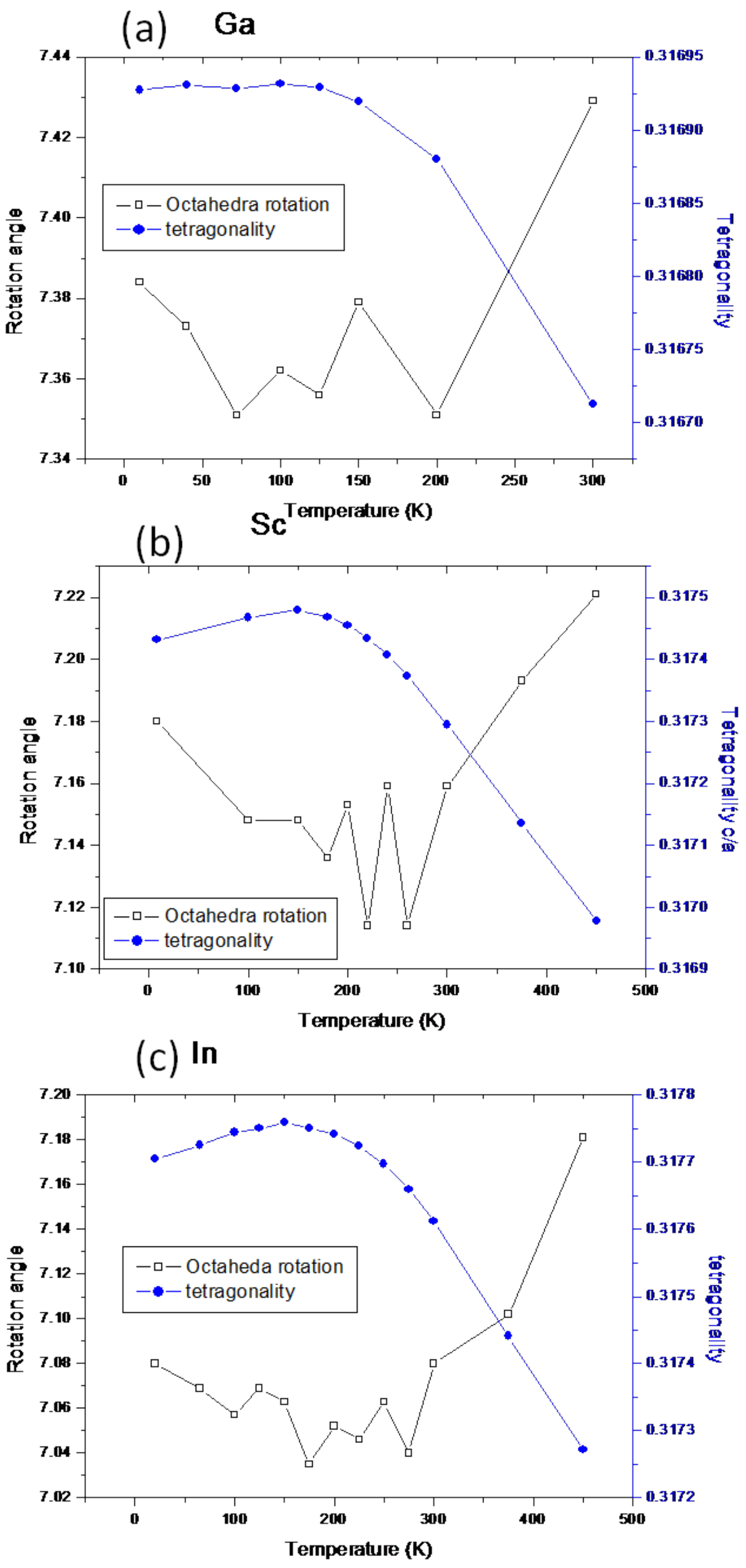

**Figure S3:** Variation of the B2 octahedral tilt angle with temperature as calculated using the O2 atomic positions for (a) $Ba_6GaNb_9O_{30}$, (b) $Ba_6ScNb_9O_{30}$ and (c) $Ba_6InNb_9O_{30}$.

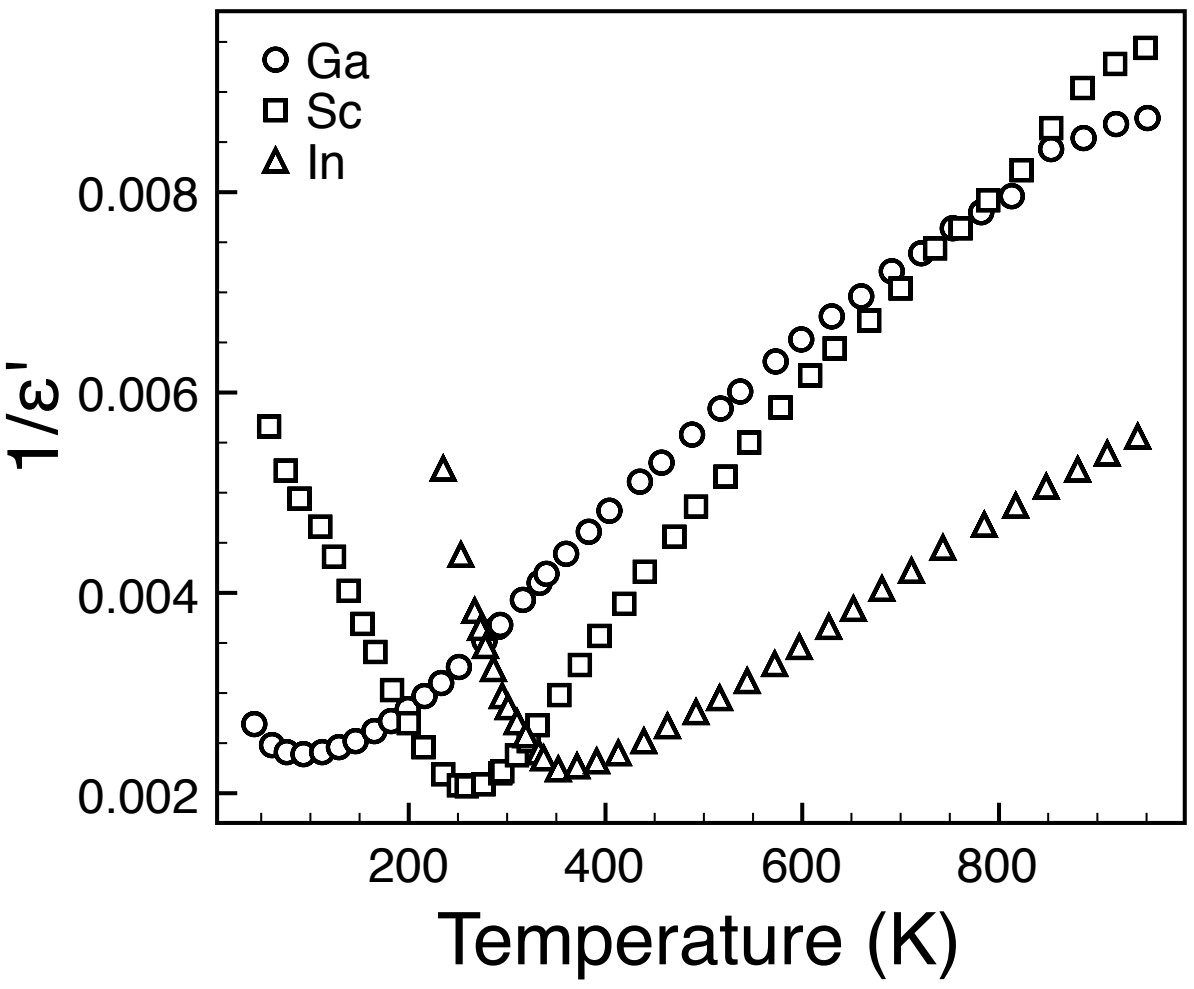

**Figure S4:** Curie-Weiss plots for $Ba_6GaNb_9O_{30}$, $Ba_6ScNb_9O_{30}$ and $Ba_6InNb_9O_{30}$. Data is at 1 MHz.